\documentclass[12pt,preprint]{aastex}

\shorttitle{Faint Blue Objects in the HDFS}
\shortauthors{Kilic et al.}

\begin{document}

\title{Faint Blue Objects in the Hubble Deep Field South Revealed: White Dwarfs, Subdwarfs, and Quasars\footnote{Based on observations made with the NASA/ESA Hubble Space Telescope, obtained from the Data Archive at the Space Telescope Science Institute, which is operated by the Association of Universities for Research in Astronomy, Inc., under NASA contract NAS 5-26555.}}

\author{Mukremin Kilic\altaffilmark{2,4}, R. A. Mendez\altaffilmark{3}, Ted von Hippel\altaffilmark{2}, and D. E. Winget\altaffilmark{2}}

\altaffiltext{2}{The University of Texas at Austin, Department of Astronomy, 1 University Station
C1400, Austin, TX 78712, USA}
\altaffiltext{3}{Departamento de Astronomia, Universidad de Chile, Casilla 36-D, Santiago, Chile}
\altaffiltext{4}{kilic@astro.as.utexas.edu}

\begin{abstract}

We explore the nature of the faint blue objects in the Hubble Deep Field South. We have derived proper motions
for the point sources in the Hubble Deep Field South using a 3-year baseline.
Combining our proper motion measurements with spectral energy distribution
fitting enabled us to identify 4 quasars and 42 stars, including 3 white dwarf candidates.  
Two of these white dwarf candidates, HDFS 1444 and 895, are found to display significant proper motion,
21.1 $\pm$ 7.9 mas yr$^{-1}$ and 34.9 $\pm$ 8.0 mas yr$^{-1}$, and are consistent with being thick disk or halo
white dwarfs located at $\sim$2 kpc. The other faint blue objects analyzed by Mendez \& Minniti do not show any
significant proper motion and are inconsistent with being halo white dwarfs; they do not contribute to the
Galactic dark matter. The observed population of stars and white dwarfs is consistent with standard Galactic models.

\end{abstract}

\keywords{dark matter --- Galaxy: halo --- stars: evolution --- white dwarfs}

\section{Introduction}

Faint blue objects discovered in deep Hubble Space Telescope images have been the subject of discussion in recent years.
The extreme depth of the Hubble Deep Field (HDF) North (Williams et al. 1996) and South (Casertano et al.
2000) and the Hubble Ultra Deep Field (HUDF; Beckwith et al. 2004) enables
us to study faint
stellar objects in the regions of the color-magnitude diagram that are devoid of standard Galactic stars.
Supported by the observed microlensing events toward the Large Magellanic Cloud (Alcock et al. 2000), several investigators
proposed that the faint blue objects observed in the HDF images can explain part of the dark matter in the Galaxy and
a significant population of the halo of the Galaxy may be in the form of low-luminosity white dwarfs. Claims by
Mendez \& Minniti (2000; hereafter M\&M) that the faint blue sources in the HDF North and South are Galactic stars seemed to be
consistent with earlier findings of Ibata et al. (1999) who found 5 faint halo white dwarf candidates with detectable proper motions in
the HDF North. However, further analysis by Richer et al. (2001) and Kilic et al. (2004) showed that the faint blue
objects in the HDF North do not show any significant proper motion. 

A detailed analysis of the point sources in the HDF North (Kilic et al. 2004) and the HUDF (Pirzkal et al. 2005)
showed that blue extra-galactic sources may be confused with white dwarfs. 
The (10$\sigma$) limiting AB magnitudes of the $I$ band images for these two fields are 27.6 and 29.0, respectively.
Using a 7 year baseline, Kilic et al. (2004) obtained proper motion measurements for the point sources in the
HDF North including the 5 faint blue objects; they identified two possible disk white dwarfs, one of which now also appears
spectroscopically to be a white dwarf (D. Stern, private communication).
Using low resolution spectroscopy and proper motion measurements, Pirzkal et al. (2005) identified 20 late type stars,
2 quasars, and 4 possible white dwarf candidates in the HUDF. Kilic, von Hippel \& Winget (2005)
showed that only two of these candidates (HUDF 4839 and 9020) are firm white dwarf candidates with HUDF 4839 possibly being
a thick disk object, and HUDF 9020 being either a disk or a halo object. None of these white dwarf candidates show significant
proper motion (Pirzkal et al. 2005). Non-detection of high velocity white dwarfs in the HUDF is consistent with
non-detection of halo white dwarfs in the HDF North. Lack of detection of high
velocity white dwarfs in these two fields implies that white dwarfs account for less than 10\% 
of the Galactic dark matter (Pirzkal et al. 2005; Kilic, von Hippel \& Winget 2005).

Ibata et al.'s (2000) and Oppenheimer et al.'s (2001) discoveries of apparent halo white
dwarfs from kinematic surveys were enough to explain 2\% of the dark matter in the solar neighbourhood.
On the other hand, further analysis by several investigators showed that most of these white dwarfs are associated with the
thick disk population of the Galaxy (Reid et al. 2001; Reyle et al. 2001; Bergeron et al. 2005).
Nevertheless, old halo white dwarfs are observed in the Globular clusters M4 (Hansen et al. 2004)
and NGC 6397 (Mendez 2002) and in the field toward M4 (Kalirai et al. 2004).
Reid (2004; see for a complete review on high velocity white dwarfs) found that these observations
do not require additions to the standard Galactic populations.

The HDF South data provides another opportunity to test whether the faint blue objects in deep Hubble images
can be old halo white dwarfs and if they can explain part of the Galactic dark matter. 
M\&M found 22 Galactic stars and 10 faint blue objects in the HDF South. If these 10 faint blue objects are halo
white dwarfs, then they would explain 30--50\% of the dark matter in the solar neighborhood.
We extend Kilic et al.'s (2004) work to the HDF South by using the original HDF South data and images of the same field
taken 3 years later
for the GO-9267 proposal (WFPC2 Supernova Search, PI: S. Beckwith) to measure the proper motions of the point sources
in the HDF South. Section 2 describes our first epoch data and the classification of the point sources, while 
data reduction procedures for our second epoch images are discussed in \S 3. \S 4 describes the proper motion
measurements for the point sources. We present our spectral energy distribution fitting results in \S 5,
and various implications of these results are then discussed in \S 6.

\section{Description of the Data}

The HDF South was imaged using the Wide Field Planetary Camera 2 (WFPC2) and the F300W (U$_{300}$), F450W (B$_{450}$), F606W (V$_{606}$), and F814W (I$_{814}$) filters in 1998, and imaged again in 2001 using the WFPC2 and the F814W filter in order to
search for high redshift supernova. The original images taken in 1998 reach down to AB = 26.8, 27.7, 28.3, and 27.7
in the four bands, respectively (10$\sigma$, point source; Casertano et al. 2000). The (version 2) source catalogs
for the first epoch are produced by the Space Telescope Science Institute (STScI) from the combined and drizzled images. 
Source detection was carried out on the inverse-variance–-weighted sum of the F606W and F814W drizzled images.
The combined F606W + F814W (hereafter $V+I$) image is significantly deeper than any of the combined images (Casertano
et al. 2000).

\subsection{Identification of Point Sources}

M\&M used the stellarity index from Source Extractor (SExtractor; Bertin \& Arnout 1996) to identify point sources and
claimed that the star/galaxy separation is reliable for objects brighter than $V+I=$29. They visually inspected the objects
with stellarity index $<$0.85 and found that they are extended. M\&M identified all objects with stellarity index $>$0.90
as stars, and found 98 point sources in 4.062 arcmin$^2$ of the HDF South. 
They analyzed objects with $I<27$ (15$\sigma$ detections) and found 22 late type stars and 10 faint blue objects
(see M\&M for a complete discussion).

Kilic, von Hippel, and Winget (2005) showed that the SExtractor stellarity index fails at faint magnitudes. Instead, they
suggested the use of the half light radius (R50) and the full width at half maximum (FWHM) measurements
along with the empirical point spread function (PSF) distributions to identify the unresolved objects
in deep HST images. Figure 1 shows the distribution of R50 with $V+I$ magnitude for all HDFS objects (top panel) and for
the objects with the stellarity index $>$0.85. Objects brighter than $V+I=20$ are saturated and therefore have higher R50 and
lower stellarity values. It is clear from this figure that the point sources are well separated from the
extended objects in this diagram down to $V+I=$28.5. A solid line (R50=3.0) marks the separation between the
resolved and the unresolved objects. We note that we trimmed the source catalog to avoid spurious detections near the
detector boundaries, and we also exclude the Planetary Camera (PC) observations because the PC has a different plate scale
and the PSFs are different from the PSFs for the Wide Field Cameras.

We use the IRAF task PRADPROF and the object centroids from Source Extractor to plot the radial profile of each object.
Our empirical PSF distributions for objects with $R50<3.0$, the stellarity index $>0.85$, and $21<V+I<28.5$
are shown in Figure 2. As in M\&M, we limit our analysis to the objects brighter than $I=27$.
The object number, R50, FWHM, and stellarity index of each object are shown in the top right corner of each
panel. We also plot the PSF for a bright unsaturated, unresolved source (HDFS 10617) in each panel for a direct comparison.
In addition, we examined the PSF distributions for the objects with $R50<3.0$ and the stellarity index $<0.85$, and (as in
M\&M) found all of them to be resolved.

A comparison of the PSF for each object with our template PSF (for HDFS 10617) shows that HDFS 126 (FWHM=6.03), 2003
(FWHM=5.38), and 497 (FWHM=5.28) are clearly resolved. The typical FWHM for the unresolved objects is $<5$. In addition,
HDFS 1812, 1827, 2129, and 567 have shallower PSF distributions than the template PSF, therefore are resolved, as well. 
Six of these seven objects have stellarity indices larger than 0.9, therefore would be classified as point sources based
on the stellarity index. Our empirical PSF distribution analysis shows that the stellarity index from the SExtractor
should not be used by itself to identify point sources in deep HST images. Instead, as Kilic, von Hippel, \& Winget (2005)
suggested, a combination of the R50, FWHM, stellarity index, and empirical PSF distributions is required to
identify unresolved objects.
This analysis leaves us with 39 unresolved objects with $21<V+I<28.5$, and $I<27$, plus 7 brighter point sources ($V+I<21$)
in the HDF South.

We note that we may have a few barely resolved objects in our sample, e.g. HDFS 191.
Some extra galactic sources, e.g. Luminous Compact Blue Galaxies (LCBGs), can be quite small (half light radius$=$1--3 kpc) and blue
($B-V<0.6$; Werk et al. 2004).
Nevertheless, HST is able to resolve almost every galaxy if enough signal to noise is obtained.
For example, Hoyos et al. (2004) could resolve four LCBGs at intermediate redshifts (z=0.1--0.44, $R50=$ 0.7--2.7 kpc) using WFPC2 ($R50=$ 0.23 -- 0.47
$\arcsec$). In addition, using low resolution spectroscopy from the GRAPES survey,
Pirzkal et al. (2005) showed that none of the unresolved objects in the HUDF ($I<27$) are LCBGs. Therefore, the number of compact
galaxies that are classified as unresolved should be small. Even if our sample is contaminated by a few faint resolved objects,
misclassification of these objects would not change the conclusions derived from our analysis (see \S 7).

Figure 3 shows the distribution of the R50 and the FWHM for the HDF South objects in different magnitude ranges. Objects with
stellarity index $>0.85$ are shown as filled triangles, and the unresolved objects identified from their PSF distributions
are shown as filled circles. All of the unresolved objects ($V+I>21$) have stellarity index $>0.87$, $R50<3$, and $FWHM<5$. 
It is clear from this figure that the unresolved objects can be identified easily down to $V+I=27$. Even though, the
classification becomes harder for $V+I>27$, five of the six objects that are fainter than $V+I=27$ and classified as unresolved
(HDFS 261, 441, 1020, 1306, and 2178) all have PSFs, R50, FWHM, and the stellarity values consistent with unresolved objects.
Therefore, their identification as point sources is secure. The identification of HDFS 2488 ($R50=$2.96, $I=$26.97) as unresolved is
questionable, but still included in our analysis for completeness purposes. Astrometric and photometric data for the
46 objects that we identified as unresolved are given in Table 1. We adopted the calibrated photometry of Casertano
et al. (2000). 

\section{The Second Epoch Data}
The second epoch data consist of thirty six 1200 sec images in the F814W filter, and provide a baseline of 3 years. The
10 $\sigma$ limiting magnitude for the point sources in the combined second epoch image is AB = 26.7.

\subsection{Data Reduction}

We followed the reduction steps outlined in the HST dither handbook (Koekemoer et al. 2002) to reduce the second epoch data.
All 36 images were processed with a procedure that included the following steps: initial pipeline processing,
sky background subtraction, cross-correlation and finding the offsets between images,
cosmic-ray rejection, scattered-light correction, and final drizzling and combination.

We obtained the standard pipeline processed images of the HDFS 2001 observations from the Space Telescope Science
Institute website. 
After subtracting the sky background from each image, cosmic rays were removed so that
cross-correlation would not be affected by them. Each image is cross correlated with the first
image (reference image) in order to determine the shifts between them. Most hot pixels, i.e.,
pixels with elevated dark current, can be identified easily with images taken in different
dither positions. Nevertheless, some hot pixels may escape detection if they fall near an
object. We use the static pixel mask reference file, used in pipeline calibration, to create
a bad pixel mask for our images.  

The cosmic-ray rejection is done in three steps. Each individual input image is registered and
drizzled to the same output frame.
Ten of the 36 images have a higher background and suffer from a distinctive cross-shaped pattern due to the shadowing of the HST
secondary by the support structure of the WFPC2 repeater (Casertano et al. 2000).
The 26 images without the scattered light problem are combined into a single cosmic-ray-free
median image for each chip. The median image is then blotted back (reverse drizzling) to the
original position of each input image. Images with the cross pattern were corrected by applying
a median filter to the image obtained after subtracting the expected image produced by the BLOT task.
Inspection of the resulting images showed that the subtraction works well.
In additon, each image is compared with the blotted images to identify and mask cosmic rays and
bad pixels.

The final drizzled images are constructed by applying the shifts and the mask files to the individual
images, and are combined into a single image for each WFPC2 chip.
The parameters chosen for the final drizzling were similar to those used for the HDFS 1998 images.
We used a footprint area of 0.6 input pixels (0.5 for the HDFS 1998 images), and a pixel scale of
0.4 pixels, resulting in a linear scale of 0.04 $\arcsec$ pixel $^{-1}$. Our experimentation with different values of the footprint for the drizzling algorithm
showed that image statistics are better for a footprint of 0.6 pixels.
Because of the relatively small pointing shifts, we decided not to combine all images into a
single mosaic comprising all four WFPC2 chips. The pointing shifts are not large enough to
recover part of the sky that is lost between the WFPC2 chips.

\subsection{Object Identification}

Casertano et al. (2000) used the SExtractor package to create source catalogs
for the HDF South 1998 images. We have created source catalogs for the second epoch data
using SExtractor, version 2.3, with the same parameters used for the HDF South original images.
Our main incentive for using SExtractor was its incorporation of weight maps in modulating the
source detection thresholds. 
SExtractor follows a standard connected pixel algorithm for source identification. We set the source
detection threshold to 0.65 and the minimum area to 16 drizzled pixels. We convolve the image
with a Gaussian (FWHM = 3 pixels) and sources with 16 pixels above the detection threshold are
included in our catalog. 
Only those objects matching the positions of the objects in the first-epoch data with differences
less than 1 pixel (0.04$\arcsec$) are included in our final catalog. Furthermore, we visually
inspected all of these sources to avoid any mismatches. We note that the 1 pixel limit is only used for the reference
compact objects, as we do not impose any limits on the proper motion of the point sources.

\section{Proper Motion Measurements}

The original HDF South images were corrected for distortion by Casertano et al. (2000).
The drizzling procedure that we used to reduce our second epoch data also corrects for distortion with
standard procedures outlined in the HST dither handbook.
Even with these distortion corrections, however, some distortion remains (Bedin et al. 2003).
Kilic et al. (2004) and Pirzkal et al. (2005) showed that a simple quadratic transformation can be
used to improve the correction for distortion in the WFPC2 and the ACS images, respectively.

To select reference objects, we use all of the compact objects (isolated, low residuals, and not fuzzy)
with positional differences
less than 1 pixel (0.04$\arcsec$) between the two epochs to derive a quadratic (third order two dimensional polynomial)
transformation for each chip (WF2, WF3, and WF4). We rejected deviant points using a 3$\sigma$ rejection
algorithm. This rejection is required because our reference objects are compact galaxies and
centroiding errors are larger for galaxies. Even though using a local tranformation for each object
can increase the accuracy of our proper motion measurements, some of the point sources are near the
edges of the chip, therefore it is not possible to perform a local transformation for each object.
Our transformation solutions for individual chips are still better than doing a global transformation
using a single mosaic image.
After mapping the distortions with the GEOMAP package, coordinates for the compact objects and the
point sources were transformed to the second-epoch positions with the GEOXYTRAN routine.

Figures 4, 5, and 6 show the observed shifts (scaled by a factor of 50) of each object on WF2, WF3,
and WF4, respectively. Crosses represent reference compact objects and filled circles are point
sources identified in the previous section. The transformation error bars are shown in the right
panel of each figure. It is clear from these figures that the reference objects do not show
any systematic motion, and several point sources are
statistically well separated from the reference objects; they are moving with respect to this
external reference frame. 

\subsection{Results}

Our proper motion measurements for the point sources in the HDF South and their significances ($\mu$/$\sigma$)
are given in Table 2. Typical errors in our measurements are 7-8 mas yr$^{-1}$. Hence, only those
objects having proper motions larger than 15 mas yr$^{-1}$ have significance greater than 2$\sigma$. Our errors in
proper motion measurements are larger than that of Kilic et al. (2004) due to the fact that the time
difference between the first and second epoch images is only 3 years versus 7 years in that study.
There are 9 point sources with $\mu$/$\sigma$$\geq$2 in our sample, and four of these stars (HDFS 10081, 10617, 2072,
and 895) display significant ($\mu$/$\sigma$$\geq$3) proper motion.

Figure 7 shows the observed shifts of all point sources along with the Galactic coordinates. Halo objects
rotate around the Galactic center more slowly than disk stars like the Sun. Solar motion
relative to the halo objects corresponds to a speed of $\sim$200 km s$^{-1}$ in the direction ${\it l}=90^{o},
{\it b}=0^{o}$ (Mihalas \& Binney 1981). Therefore, halo objects are expected to lag behind the disk objects in the opposite
direction to the rotation direction of the Sun (${\it l}$). A comparison of Figures 4, 5, 6, and 7 shows that even though the
reference compact objects show a symmetric distribution around the origin, most of the point sources in the HDF South are
located in the lower left quadrant of the figure; the majority of them lag behind the Sun, therefore are likely to be
halo objects.

Halo stars are expected to have large proper motions as a result of their high velocities relative
to the Sun. M\&M claimed that the faint blue objects in the HDF South could be
ancient halo white dwarfs with distances, based on consistency with white dwarf absolute magnitudes, less than about 2 kpc
from the Sun. Assuming $V_{\rm tan}=$ 200
km s$^{-1}$ for a halo star at a distance $\leq$ 2 kpc, we would expect to measure a proper motion of
$\geq$ 21 mas yr$^{-1}$ for the faint blue objects.
Nine of the faint blue objects identified by M\&M (HDFS 1812, 1827,
1945, 2007, 2178, 441, 1020, 261, 1332) have proper motions in the range 2.26 to 10.27 mas
yr$^{-1}$, therefore are not likely to be halo white dwarfs. The remaining object, HDFS 1444, exhibits
a significant proper motion, 21.07 $\pm$ 7.93 mas yr$^{-1}$, and is likely to be either a thick disk or a halo
object. If the majority of these faint blue objects are not halo white dwarfs, what are they?

\section{Spectral Energy Distribution Fitting}

It is always desirable to obtain spectroscopy of star-like objects to distinguish quasars from stars, and
also to determine the spectral types of stars in order to have a reliable estimate of their
absolute magnitudes and kinematic properties. Unfortunately, the most interesting point sources in deep
Hubble images are the faintest, therefore the hardest ones for which to obtain spectroscopy. There is only one
deep field, the Hubble Ultra Deep Field, that is studied spectroscopically to very faint magnitudes
(I$\leq$27; Pirzkal et al. 2005). Pirzkal et al. (2005) found 28 point sources in the Hubble Ultra Deep
Field, including 2 quasars (and additional two more quasars with I$>$27). Even though
spectroscopy is superior to spectral energy distribution fitting using photometry, Kilic, von Hippel, \& Winget (2005)
could identify the quasars and the stars in the HUDF from their photometry. Using the IRAF package CALCPHOT, which
is designed for simulating the HST observations, Kilic, von Hippel, \& Winget (2005) showed that spectral energy
distribution (SED) fitting can be used to identify quasars, and to assign spectral types to stars.
Their classifications for the stars agreed reasonably well with the Pirzkal et al. (2005) spectroscopic classifications,
and they were also able to predict the redshifts of two of the quasars correctly, whereas the other two
were predicted to be at lower redshifts than the spectroscopic redshift measurements. 

Here we adopt the same SED fitting procedures used by Kilic, von Hippel, \& Winget (2005).
We use Pickles (1998) stellar templates and the composite-quasar spectrum from the Sloan Digital Sky Survey
(Vanden Berk et al. 2001) to simulate SEDs in U$_{300}$, B$_{450}$, V$_{606}$, and I$_{814}$ using the CALCPHOT task.
Our simulations include stellar templates from O5 to M6 dwarf stars, and the composite quasar spectrum is
used to simulate the colors for quasars up to $z=4.8$.
Note that we do not simulate the colors of the broad range of galaxy types, as almost every galaxy with sufficient
signal-to-noise is resolved by HST (e.g. 2 kpc at $z$ = 0.5 projects to 0.33$\arcsec$ for
$\Omega_{\rm o}$ = 0.3, $\Lambda_{\rm o}$ = 0.7, and H$_{\rm o}$ = 70).

The observed magnitudes for the faint blue objects are converted to $F_{\rm \nu}$, normalized at 
V$_{606}$, and compared with our simulated SEDs. We assign spectral types to each object using a
$\chi^2$ minimization technique. Each photometry point is weighted according to its errorbar, but we also tried giving
equal weights to each photometric band to explore possible fits. Figure 8 shows our best-fit solutions to the point
sources in the HDF South. Observed fluxes are shown as filled circles, the solid lines show the best fit stellar or
quasar templates
when each photometric band is weighted according to its errorbar, and the dashed lines show the best fit templates assuming
equal weights for all bands. The object number and the assigned spectral types are given in the top left corner of
each figure. Error bars for the observed fluxes are so small that they can only be seen
for the fainter objects. We note that the $U$ and $B$ photometry for HDFS 2615 is probably wrong, as it is barely detected
in the $U$ and $B$ images, yet its photometry shows excesses in the $U$ and the $B$ filters, also apparent in Figure 13.
Even though our SED fitting algorithm gives a range of spectral types (O9-M2) for this object, it is more likely to be
an M type star based on more accurate $V$ and $I$ photometry.

Figure 8 shows that there are 4 point-like sources with SEDs better explained by quasars than stars.
HDFS 10151 ($z=2.0$), 1945 ($z=0.2$), 2007 ($z=4.7$), and 2178 ($z=4.0$) have colors more consistent
with being quasars. Interestingly enough, three of these objects were classified as possible halo
white dwarf candidates by M\&M. These four objects have proper motions in the range
2.56 -- 4.91 $\pm$ 8 mas yr$^{-1}$ (see Table 2), and are not moving (within the errors); they demonstrate that
our proper motion measurements are reliable.

\section{Discussion}

We identified 4 quasars and 42 stars, including 36 K0 or later type stars, with our SED fitting technique. We adopt
the absolute magnitude ($M_{\rm V}$) for each spectral type from Pickles (1998; Table 2). We use the $M_{\rm V}$
for each object to simulate the absolute magnitudes in the WFPC2 F606W filter ($M_{\rm F606W}$). 
We calculated photometric distances for all of the objects in our sample using the simulated
$M_{\rm F606W}$ and the observed $V_{\rm 606}$ magnitudes. Combining our proper motion measurements with photometric
distances, we are able to derive tangential velocities for the stars.
The photometric errors for the point sources in the HDF South are small, hence they cause only relatively small errors in
the estimated distances. On the other hand, if the assigned spectral types are wrong by 1 index (for example
M3 instead of M2), then the absolute magnitudes could be wrong by as much as 1 magnitude.
We also note that our stellar templates have approximately solar metallicity. Therefore, the distances to the metal
poor halo objects, which will be intrinsically fainter for the same spectral type, may be over-estimated by our SED
fitting method. For example, a metal poor G0 dwarf ($[Fe/H]=-0.8$) would be 0.5 magnitude fainter than a solar metallicity G0
star (Pickles 1998). Hence, the distance to a metal poor G0 type star would be over-estimated by 26\%.

In order to determine the effect of different metallicites, we use synthetic spectra from a PHOENIX model atmosphere grid
(Brott \& Hauschildt 2005) for stars with 2000 K $\leq T_{\rm eff} \leq$ 10000 K, [Z/H]= 0.0, -0.5,-2.0, and log(g)=4.5 to
simulate photometric colors in the HST bands. We found that using different metallicities changes the best fit
$T_{\rm eff}$ by 50 -- 450 K for G0 and later type stars, and the spectral types obtained from metal poor atmosphere models
are usually 1-2 later in spectral types, e.g. M2 instead of M0, than the ones obtained from the models with solar metallicity.
We also used Kurucz model atmospheres (Kurucz 1995) with 3500 K $\leq T_{\rm eff} \leq$ 6000 K to test the metallicity
effect, and found similar results. Using Girardi et al. (2002) theoretical isochrones,
we estimate $M_{\rm V}$ and $M_{\rm F606W}$ for different metallicities for the point sources in our sample.
Spectral types from the SED fitting
procedure (using the Pickles library), estimated absolute magnitudes, derived distances, and tangential velocities are given
in Table 3. The ranges of absolute magnitudes, distances, and tangential velocities show the effect of using models with
different metallicities. 
Despite these potential systematic errors, the total information we have available for the point sources (the morphological
information, proper motion measurements, and the SED fitting results) is sufficient to determine their nature. 

\subsection{Late Type Stars in the Galaxy}
Figure 9 shows the histogram of the number of stars observed at a given distance in the HDF South along with the predictions
from the star count models (Reid \& Majewski 1993; dashed line) and the observed distribution of stars in the HUDF
(Pirzkal et al. 2005; solid line). The top panel shows the distribution of stars
if we use Pickles main sequence library, and the bottom panel shows the same distribution if we use synthetic spectra
with [Z/H]=-2.0 (halo-like metallicity). 
Both panels show that there is a deficit of stars in the 5--15 kpc range in the HDF South compared to the star count models.

If we assume that all of the stars in the HDF South have solar metallicity, then we see an excess number of stars at
$\sim$35 kpc, and 4 more stars (HDFS 441, 1020, 2596, and 261) with estimated distances larger than 50 kpc,
which is not expected from the star count models, nor would the distances be consistent with the metallicities for Galactic
stars. Most of the objects observed at $\sim$35 kpc are K type stars, whereas
the four objects with distances $>50$ kpc fit late type star SEDs fairly well, and they have
essentially the same spectral type (K7-M0). If we assume that their classification as point sources is reliable,
our SED fitting procedure works well to identify quasars and stars, and that they have solar metallicity,
then we could claim that we discovered two new populations of stars; a cluster of stars at 35 kpc and several other
stars at $>$ 50 kpc.

HDF South and HUDF have similar Galactic latitudes (-49.21 and -54.39, respectively), therefore they
should have similar distributions of Galactic objects. We do not find any stars with distances
larger than 50 kpc in the HDF North or the HUDF. Even though HDF South ($l=$328.25, $b=$-49.21) is
about 25$^{o}$ away from the center of the Small Magellanic Cloud (SMC; $l=$302.80, $b=$-44.31), the Magellanic
Stream has over dense regions near the HDF South (1$\times$ 10$^{19}$ atoms cm$^{-2}$; see the HI surface density maps
of Mathewson \& Ford 1984). The average distance to the 4 objects with $d>50$ kpc is 66.8 $\pm$ 13.4 kpc (assuming solar
metallicity), whereas the distance to the SMC is measured to be 60.6 $\pm$ 3.8 kpc (Hilditch, Howarth, \& Harries 2005).
Can these objects be members of the SMC? Assuming that they have metallicities similar to the SMC ($[Fe/H]=-0.7$;
Lennon 1999), we estimate their average distance to be 17.8 $\pm$ 3.4 kpc, therefore they are not consistent with
being in the SMC.
                                        
We expect the majority of the stars in the HDF South to be halo stars (from the star count models and the velocity
distribution of Figure 7). Therefore, the bottom
panel of Figure 9 is likely to represent the real distribution of stars better than the top panel. 
A comparison of the distances, tangential velocities, and photometric colors show that
all of the 42 stars except three (HDFS 1444, 895, and 2488) are consistent with being dwarf/subdwarf stars in the Galaxy.
Seven of these 42 stars (HDFS 191, 1576, 1922, 2072, 10081, 10326, and 10617) show
significant proper motion ($\mu/\sigma\geq2$) and have spectral types M2--M5, therefore are
probable halo M dwarfs. Large errors in our proper motion measurements prevent us from classifying
the kinematic properties of the other stars, nevertheless, they are most likely to be G0 and later type dwarfs
in the thick disk or halo of the Galaxy.
The three stars (HDFS 1444, 895, and 2488) with estimated distances larger than 90 kpc (both for the metal rich and
the metal poor case) are discussed in the next section.

\subsection{White Dwarfs}
HDFS 1444, 895, and 2488 would have to be at very large distances ($>90$ kpc) if they were main-sequence stars of any
metallicity. On the other hand, as white dwarfs, they would be at more reasonable distances.
In order to find the temperatures of these objects, we simulated the colors for blackbody SEDs with
temperatures in the range 3000 -- 80000 K.
Figure 10 shows the best fitting blackbody SEDs (solid lines) for HDFS 1444 (top panel), 895 (middle panel), and HDFS 2488
(bottom panel). We calculate the blackbody temperatures for these objects to be 10547 K, 6096 K, and 10463 K,
respectively. We have also used DA white dwarf models (D. Saumon and D. Koester, private communication) to simulate colors for
white dwarfs with $log$ g = 8 and 3000 K $\leq T_{\rm eff}
\leq$ 20000 K. Cool white dwarfs show depressed infrared fluxes due to the effects of collision induced absorption (CIA)
due to molecular hydrogen (Hansen 1998, Saumon \& Jacobson 1999). The pure H white dwarf models that we used include the CIA
opacities, therefore, we are able to compare the spectral energy distributions of young and old white dwarfs simultaneously
and find the best-fit solution for our white dwarf candidates.
Assuming that HDFS 1444, 895, and 2488 are pure-H atmosphere white dwarfs, we estimate the temperatures of these
objects to be 10681 K, 5882 K, and 11000 K, respectively.
Our best fitting DA white dwarf SEDs are shown as dotted lines in Figure 10.

Using our best fit DA white dwarf atmosphere solutions, we estimate the absolute magnitudes for our white dwarf candidates
using the tables from Bergeron et al. (1995). We use the white dwarf models
to predict $M_{\rm F606W}$, and therefore to calculate the distances and tangential velocities for these three
objects. Estimated distances and kinematic properties of our white dwarf candidates
are given in Table 3. HDFS 1444 displays a significant proper motion, 21.07 $\pm$ 7.93 mas yr $^{-1}$,
and is consistent with being a thick disk or halo object at 1.5 kpc with $V_{\rm tan}=$ 148 $\pm$ 56 km s$^{-1}$.
Likewise, HDFS 895 displays a proper motion of 34.9 $\pm$ 8.0 mas yr $^{-1}$, and is more likely to be a halo white
dwarf at 2.1 kpc with $V_{\rm tan}=$ 346 $\pm$ 79 km s$^{-1}$.
HDFS 2488 does not display any significant proper motion, and its classification as a point source is questionable (see
\S 2.1), nevertheless, if it is a star, then it would have to be a halo white dwarf at 9 kpc.

Figure 11 shows the $U$ vs. $V$ velocity diagram for the stars (triangles; assuming [M/H]=-2.0) and the likely white dwarfs
(circles) in the HDF South. We use the results of Chiba \& Beers (2000) for the expected velocity distribution of halo
(solid line) and thick disk (dashed line) objects. It is apparent from this figure that we have several thick disk stars
in our sample. Nevertheless, the majority of the stars are more likely to be in the Galactic halo.
One of the white dwarf candidates, HDFS 1444, may well be a thick disk white dwarf, whereas the other two candidates,
HDFS 895 and 2488, are more likely to be halo white dwarfs.

\subsection{Mendez \& Minniti Faint Blue Objects}

M\&M identified 10 faint blue objects in the HDF South. We classified two of these 10 objects
(HDFS 1812 and 1827) as resolved (see Figure 2). We measure proper motions of 2.26 $\pm$ 7.93 and 3.25 $\pm$ 6.98 mas
yr$^{-1}$ for these two objects, respectively. One of the faint blue objects (HDFS 1332) is fainter than $I_{\rm 814}=27$
(therefore not included in our analysis), and has a proper motion of 9.31 $\pm$ 7.93 mas yr$^{-1}$.
Six of the faint blue objects (HDFS 1945, 2007, 2178, 441, 1020, 261) have proper motions in the range 2.56 to 10.27 mas
yr$^{-1}$. Our SED fitting analysis showed that HDFS 1945, 2007, and 2178 have colors more consistent with being quasars
than stars. In addition, we classify HDFS 441, 1020, and 261 as metal poor stars in the halo of the Galaxy.
Therefore, only one of the faint blue objects identified by M\&M, HDFS 1444, plus two more white dwarf candidates identified in
our analysis (HDFS 895 and 2488) are consistent with being white dwarfs.

We use Reid \& Majewski (1993) star count models and our own calculations based on Gilmore et al. (1989; see Kilic et al.
2004 for a complete
discussion) to predict the number of stars and white dwarfs expected in the HDF South. We expect to find 45--52 stars and
0.66--2.31 white dwarfs, including 0.24--0.50 disk white dwarfs in the HDF South. The star count models mildly over-predict
the observed number of stars. The observed population of 2--3 white dwarfs is consistent with the standard Galactic models.

\section{Conclusion}
Superb resolution of the HST WFPC2 camera helped us to eliminate almost all of the large number of galaxies present in the HDF South.
Using accurate SED fitting procedures (to eliminate quasars) and obtaining proper motion measurements with a 3 year
baseline (to identify fast moving--halo objects) enabled us to classify the point sources in the HDF South.
We identified 4 quasars (consistent with zero proper motion) and 42 stars. Three of these stars (HDFS 1444, 895, and
2488), if on the main sequence,
are too distant to be in the Galaxy, and are best explained as white dwarf stars. Their kinematic properties show that
HDFS 1444 is probably a thick disk object, whereas the other two are more likely to be in the halo of the Galaxy. Third
epoch data on the HDF South would be useful to place better constraints on the proper motions and the kinematic memberships
of the point sources.

Figure 12 and 13 show the location of the stars, the quasars, and the white dwarf candidates in color-magnitude and
color-color diagrams. A comparison of Figure 12 with Figure 3 of M\&M shows that a detailed analysis of the faint
blue objects is needed to classify their nature--important additional constraints can be derived from SED fitting and from proper
motions, even with a baseline of only 3 years. The number of blue extra-galactic sources is enormous at these magnitudes
compared to the number of stellar objects. Our analysis shows that only two of the faint blue objects (HDFS 1444 and 895) show
significant proper motion and are consistent with being disk/halo white dwarfs. None of the other faint blue objects exhibit
significant proper motion, and therefore they are highly unlikely to be halo white dwarfs.
Even if we misclassified a few blue extra-galactic objects as unresolved, this would only decrease the observed number of stars and white dwarfs in the
HDF South, and therefore, strengthen our conclusion that the majority of the faint blue objects are not halo white dwarfs. 
The observed population of $\sim$2 white dwarf candidates
($I<27$) in the HDF
North (Kilic et al. 2004), HDF South (this study), and HUDF (Kilic, von Hippel, \& Winget 2005) imply that the faint blue
objects, and especially the non-observed population of halo white dwarfs, are highly unlikely to solve the dark matter
problem.

\acknowledgements

We thank Didier Saumon and Detlev Koester for kindly providing us their white dwarf models.
We are grateful to I. Neill Reid for useful discussions on calculating space velocities of stars. 
We especially thank our anonymous referee for helpful suggestions that greatly improved the article.
This material is based upon work supported by the National Science Foundation under Grant AST-0307315
to TvH, DEW, and MK. RAM acknowledges support from the Chilean Centro de Astrof\'{\i}sica FONDAP (No. 15010003).

\clearpage
\begin{deluxetable}{rlrrccccc}
\tabletypesize{\footnotesize}
\tablecolumns{9}
\tablewidth{0pt}
\tablecaption{Point Sources in the Hubble Deep Field South}
\tablehead{
\colhead{No}&
\colhead{Object}&
\colhead{X(HDFS)}&
\colhead{Y(HDFS)}&
\colhead{$U_{300}$}&
\colhead{$B_{450}$}&
\colhead{$V_{606}$}&
\colhead{$I_{814}$}& \colhead{Stellarity}}
\startdata
10663& J223250.50-603400.8& 3039.857& 884.384& 23.43& 19.90& 19.13& 18.20& 0.72\\
10692& J223257.00-603405.7& 1836.189& 772.533& 20.89& 19.14& 19.22& 18.61& 0.94\\
10017& J223254.90-603144.1& 2255.468& 4321.344& 24.99& 21.47& 20.26& 18.75& 0.78\\
10666& J223305.04-603400.8& 350.133& 907.811& 25.62& 21.80& 20.52& 19.24& 0.99\\
10151& J223250.51-603218.8& 3059.869& 3445.525& 21.76& 20.25& 20.35& 19.87& 0.99\\
10081& J223247.45-603160.0& 3632.237& 3911.988& 27.14& 23.93& 22.23& 19.89& 0.89\\
1576& J223303.65-603330.6& 614.027& 1664.433& 28.93& 23.45& 21.92& 20.23& 0.98\\
2257& J223305.63-603358.2& 242.259& 974.553& \nodata& 24.50& 22.87& 21.15& 0.99\\
10617& J223258.30-603351.7& 1599.243& 1127.375& 29.19& 25.26& 23.55& 21.36& 0.98\\
86& J223248.06-603148.6& 3521.148& 4199.046& 23.98& 22.24& 21.76& 21.46& 0.94\\
15& J223248.47-603139.3& 3447.529& 4431.511& 27.51& 24.78& 23.28& 21.59& 0.96\\
2041& J223251.03-603353.5& 2942.939& 1069.948& 28.28& 25.97& 24.16& 21.68& 0.99\\
10326& J223252.60-603259.5& 2664.651& 2425.551& 27.64& 24.45& 22.87& 21.71& 0.98\\
701& J223251.33-603237.6& 2903.785& 2973.677& 29.04& 24.80& 23.24& 21.78& 0.96\\
1922& J223247.58-603347.3& 3583.103& 1217.824& \nodata& 24.67& 23.05& 21.87& 0.88\\
1209& J223245.74-603309.7& 3931.750& 2158.995& \nodata& 26.93& 24.75& 22.24& 0.98\\
1257& J223256.68-603313.8& 1906.621& 2075.463& \nodata& 26.21& 24.56& 22.46& 0.97\\
431& J223247.94-603216.6& 3537.732& 3496.018& 25.68& 23.69& 23.14& 22.69& 0.93\\
1187& J223258.46-603307.7& 1578.305& 2231.191& 28.61& 25.28& 23.79& 22.74& 0.99\\
2469& J223253.24-603413.4& 2529.645& 574.651& 26.26& 23.89& 23.22& 22.83& 0.98\\
323& J223246.71-603207.2& 3765.987& 3728.892& 26.20& 24.07& 23.41& 22.97& 0.98\\
108& J223249.34-603149.0& 3284.156& 4189.406& 27.78& 25.54& 24.01& 22.97& 0.98\\
1444& J223258.77-603323.5& 1517.394& 1833.822& 23.30& 22.69& 22.80& 23.02& 0.97\\
2364& J223257.72-603408.7& 1702.207& 699.988& 28.80& 27.21& 25.31& 23.04& 0.98\\
1724& J223258.11-603337.4& 1637.423& 1485.485& 26.72& 24.26& 23.60& 23.12& 0.98\\
1386& J223248.36-603319.0& 3444.716& 1930.820& 28.01& 24.99& 23.86& 23.16& 0.94\\
933& J223247.94-603251.1& 3528.871& 2630.077& \nodata & 26.35& 24.75& 23.24& 0.98\\
2072& J223255.36-603355.0& 2142.223& 1039.552& 27.33& 26.90& 25.45& 23.31& 0.98\\
1331& J223247.05-603316.0& 3687.492& 2003.460& 28.08& 26.35& 24.68& 23.50& 0.98\\
316& J223253.09-603206.7& 2584.979& 3752.969& 29.53& 26.00& 24.57& 23.75& 0.98\\
1010& J223256.14-603256.9& 2011.083& 2498.703& 27.27& 24.84& 24.23& 23.78& 0.98\\
1426& J223304.54-603322.5& 450.919& 1867.914& 30.99& 26.48& 25.04& 24.10& 0.99\\
652& J223251.18-603234.0& 2933.967& 3064.594& 28.54& 26.14& 24.98& 24.30& 0.99\\
191& J223256.17-603156.5& 2018.212& 4011.988& \nodata& 28.28& 26.82& 24.41& 0.95\\
1302& J223252.60-603315.1& 2661.049& 2035.617& \nodata& 27.28& 25.58& 24.66& 0.99\\
895& J223246.22-603248.4& 3847.227& 2693.725& 28.08& 26.20& 25.77& 25.57& 0.99\\
2615& J223258.88-603421.9& 1485.890& 368.651& 25.45& 25.17& 26.91& 25.69& 0.98\\
1945& J223249.59-603347.7& 3211.154& 1212.031& 26.23& 26.53& 26.46& 25.88& 0.97\\
2007& J223249.63-603351.3& 3201.802& 1122.227& 29.13& 28.72& 26.69& 25.92& 0.99\\
2596& J223254.05-603419.4& 2378.829& 425.350& \nodata& 27.73& 26.89& 26.00& 0.99\\
441& J223255.01-603217.7& 2228.504& 3480.124& 29.78& 28.34& 27.12& 26.14& 0.98\\
1020& J223254.03-603257.4& 2400.755& 2481.265& 29.20& 28.83& 27.12& 26.29& 0.98\\
2178& J223258.94-603400.1& 1478.979& 917.399& 28.73& 28.41& 26.71& 26.47& 0.98\\
261& J223248.69-603200.8& 3402.451& 3893.844& 31.80& 28.74& 27.86& 26.91& 0.87\\
1306& J223258.20-603315.2& 1625.787& 2040.621& 29.73& \nodata& 28.47& 26.93& 0.98\\
2488& J223245.71-603413.4& 3922.570& 561.776& 28.20& 26.45& 26.65& 26.97& 0.97
\enddata
\end{deluxetable}

\clearpage
\begin{deluxetable}{rcrrrcc}
\tabletypesize{\scriptsize}
\tablecolumns{7}
\tablewidth{0pt}
\tablecaption{Proper Motions}
\tablehead{
\colhead{Object}&
\colhead{Chip}&
\colhead{$\mu_x$(pixel)}&
\colhead{$\mu_y$(pixel)}&
\colhead{$\mu$(mas/yr)}&
\colhead{$\sigma_{\mu}$(mas/yr)}&
\colhead{$\mu/\sigma$}}
\startdata
10663& wf3& -0.67& -0.58& 11.82& 7.04& 1.68\\
10692& wf2& 0.00& -0.41& 5.50& 7.93& 0.69\\
10666& wf2& -0.68& -0.30& 9.85& 7.93& 1.24\\
10151& wf4& -0.04& -0.29& 3.94& 8.00& 0.49\\
10081& wf4& -1.89& -0.84& 27.58& 8.00& 3.44\\
1576& wf2& -1.57& -0.23& 21.17& 7.93& 2.67\\
10617& wf2& -1.54& -0.94& 24.01& 7.93& 3.03\\
86& wf4& -0.11& -0.30& 4.21& 8.00& 0.53\\
2041& wf3& 0.11& 0.47& 6.47& 7.04& 0.92\\
10326& wf4& 0.23& -1.48& 20.00& 8.00& 2.50\\
701& wf4& -0.31& -1.08& 14.94& 8.00& 1.87\\
1922& wf3& 0.55& -1.20& 17.57& 7.04& 2.50\\
1257& wf2& -0.33& -0.64& 9.56& 7.93& 1.21\\
431& wf4& 0.17& 0.12& 2.79& 8.00& 0.35\\
2469& wf3& -0.29& -0.33& 5.86& 7.04& 0.83\\
323& wf4& 0.26& -0.35& 5.83& 8.00& 0.73\\
108& wf4& -0.92& -0.39& 13.33& 8.00& 1.67\\
1444& wf2& 0.05& -1.58& 21.07& 7.93& 2.66\\
2364& wf2& -0.03& -0.20& 2.70& 7.93& 0.34\\
1724& wf2& -0.10& -0.32& 4.48& 7.93& 0.56\\
1386& wf3& -0.82& -0.15& 11.07& 7.04& 1.57\\
933& wf4& -0.19& -0.44& 6.47& 8.00& 0.81\\
2072& wf3& -1.38& -0.80& 21.28& 7.04& 3.02\\
1331& wf3& -0.21& -0.55& 7.83& 7.04& 1.11\\
316& wf4& -0.10& -0.38& 5.29& 8.00& 0.66\\
1426& wf2& -0.91& -0.45& 13.50& 7.93& 1.70\\
652& wf4& -0.31& -0.36& 6.32& 8.00& 0.79\\
191& wf4& -0.98& -0.78& 16.73& 8.00& 2.09\\
1302& wf3& -0.34& -0.20& 5.22& 7.04& 0.74\\
895& wf4& -0.34& -2.60& 34.90& 8.00& 4.36\\
2615& wf2& 0.32& -0.48& 7.62& 7.93& 0.96\\
1945& wf3& 0.30& 0.21& 4.91& 7.04& 0.70\\
2007& wf3& -0.15& -0.17& 3.03& 7.04& 0.43\\
2596& wf3& -0.03& -0.14& 1.90& 7.04& 0.27\\
441& wf4& -0.75& -0.19& 10.27& 8.00& 1.28\\
1020& wf4& -0.44& -0.57& 9.63& 8.00& 1.20\\
2178& wf2& 0.14& -0.13& 2.56& 7.93& 0.32\\
261& wf4& 0.35& -0.27& 5.94& 8.00& 0.74\\
1306& wf2& -0.54& 0.09& 7.34& 7.93& 0.93\\
2488& wf3& -0.38& -0.17& 5.60& 7.04& 0.80
\enddata
\tablecomments{HDFS 15, 1010, 1187, 1209, 2257, and 10017 are out of the field of view in the second epoch data.}
\end{deluxetable}

\clearpage
\begin{deluxetable}{rcccc}
\tabletypesize{\footnotesize}
\tablecolumns{7}
\tablewidth{0pt}
\tablecaption{Spectral Types, Photometric Distances, and Tangential Velocities}
\tablehead{
\colhead{Object}&
\colhead{Type}&
\colhead{$M_{F606W}$}&
\colhead{$d$(kpc)}&
\colhead{$V_{tan}$(km/s)}}
\startdata
10663& K1-M0& 5.73--8.25& 1.5--4.8& 84--269\\
10692& A3-4& 1.42--4.97& 7.1--36.4& 189--948\\
10017& K5-6& 7.33--10.58& 0.9--3.9&\nodata\\
10666& M2& 9.17--11.09& 0.8--1.9& 36--87\\
10151& QSO& z=2.0&\nodata &\nodata\\
10081& M5& 12.18--12.46& 0.9--1.0& 118--134\\
1576& M3& 10.19--12.56& 0.7--2.2& 75--222\\
2257& M3& 10.19--12.56& 1.2--3.4&\nodata \\
10617& M5& 12.18--12.46& 1.7--1.9& 188--214\\
86& G3& 4.65--6.44& 12.0--26.4& 240--527\\
15& M3& 10.19--12.56& 1.4--4.2&\nodata \\
2041& M5& 12.18--12.46& 2.2--2.5& 67--76\\
10326& M2& 9.17--11.09& 2.3--5.5& 216--522\\
701& M2.5& 9.68--11.70& 2.0--5.2& 144--365\\
1922& M2& 9.17--11.09& 2.5--6.0& 206--498\\
1209& M6& 10.86--13.56& 1.7--6.0&\nodata \\
1257& M5& 12.18--12.46& 2.6--3.0& 119--136\\
431& K1& 5.73--7.24& 15.1--30.3& 200--400\\
1187& M1& 8.69--11.14& 3.4--10.5&\nodata\\
2469& K1& 5.73--7.21& 15.9--31.5& 442--874\\
323& K1-3& 5.73--7.43& 15.7--34.4& 434--949\\
108& M1& 8.69--11.14& 3.7--11.6& 237--731\\
1444& B8-9& -1.39 -- -0.09& 379.2--690.0& 37875--68921\\
1444& WD  & 11.95 & 1.482 & 148 $\pm$ 56\\
2364& M5& 12.18--12.46& 3.7--4.2& 48--54\\
1724& K3& 6.58--7.68& 15.2--25.3& 324--537\\
1386& K6& 7.59--10.64& 4.4--18.0& 231--942\\
933& M3& 10.19--12.00& 3.6--8.2& 109--251\\
2072& M5& 12.18--12.46& 4.0--4.5& 399--454\\
1331& M2& 9.17--11.09& 5.2--12.6& 194--469\\
316& K7& 7.86--11.15& 4.8--22.0& 121--552\\
1010& K3& 6.58--7.41& 23.1--33.9&\nodata \\
1426& M0& 8.25--11.13& 6.1--22.8& 388--1460\\
652& K6& 7.59--10.50& 7.9--30.1& 236--901\\
191& M5& 12.18--12.46& 7.4--8.5& 589--670\\
1302& M0& 8.25--11.13& 7.8--29.3& 193--725\\
895& G0& 4.18--5.98& 93.5--208.4& 15469--34471\\
895& WD& 14.17 & 2.094& 346 $\pm$ 79\\ 
2615& O9-M2& -4.32--11.09& 14.6--17600& 527--636472\\
1945& QSO& z=0.2&\nodata &\nodata \\
2007& QSO& z=4.7&\nodata &\nodata \\
2596& M0& 8.25--11.13& 14.2--53.3& 127--479\\
441& M0& 8.25--11.13& 15.7--59.3& 766--2886\\
1020& K7& 7.86--11.15& 15.6--71.1& 713--3244\\
2178& QSO& z=4.0&\nodata &\nodata \\
261& M0& 8.25--11.13& 22.2--83.5& 623--2349\\
1306& M3& 10.19--12.18& 18.1--45.4& 631--1577 \\
2488& A0-1& 0.71--1.00& 1350--1540& 35803--40918\\
2488& WD& 11.86 & 9.070 & 241 $\pm$ 303 
\enddata
\end{deluxetable}

\clearpage
\begin{figure}
\figurenum{1}
\plotone{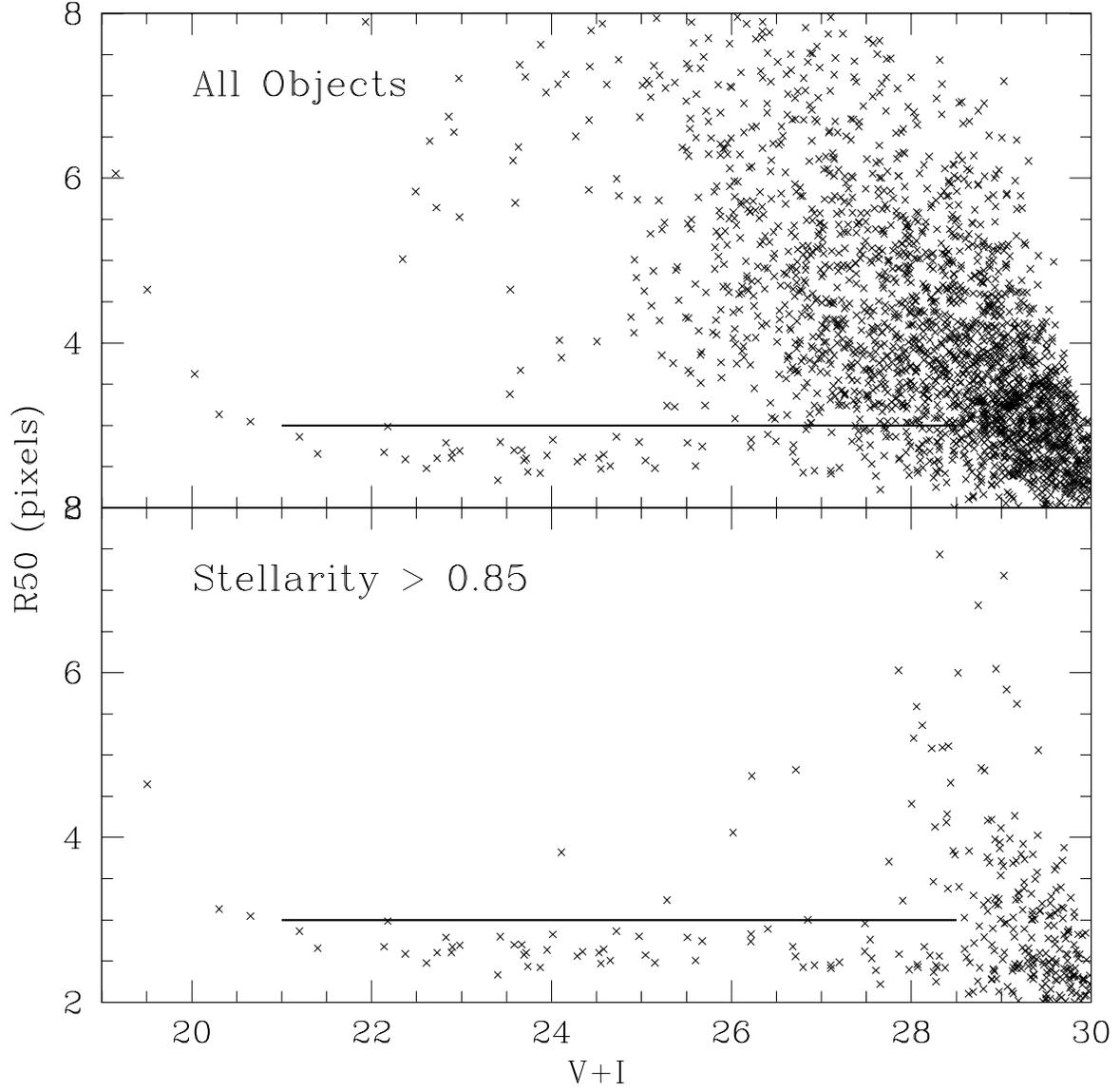}
\caption{The distribution of R50 (half light radius) for the HDF South objects. The solid line separates the
unresolved objects from the resolved objects.}
\end{figure}

\begin{figure}
\figurenum{2}
\includegraphics[angle=-90,scale=.7]{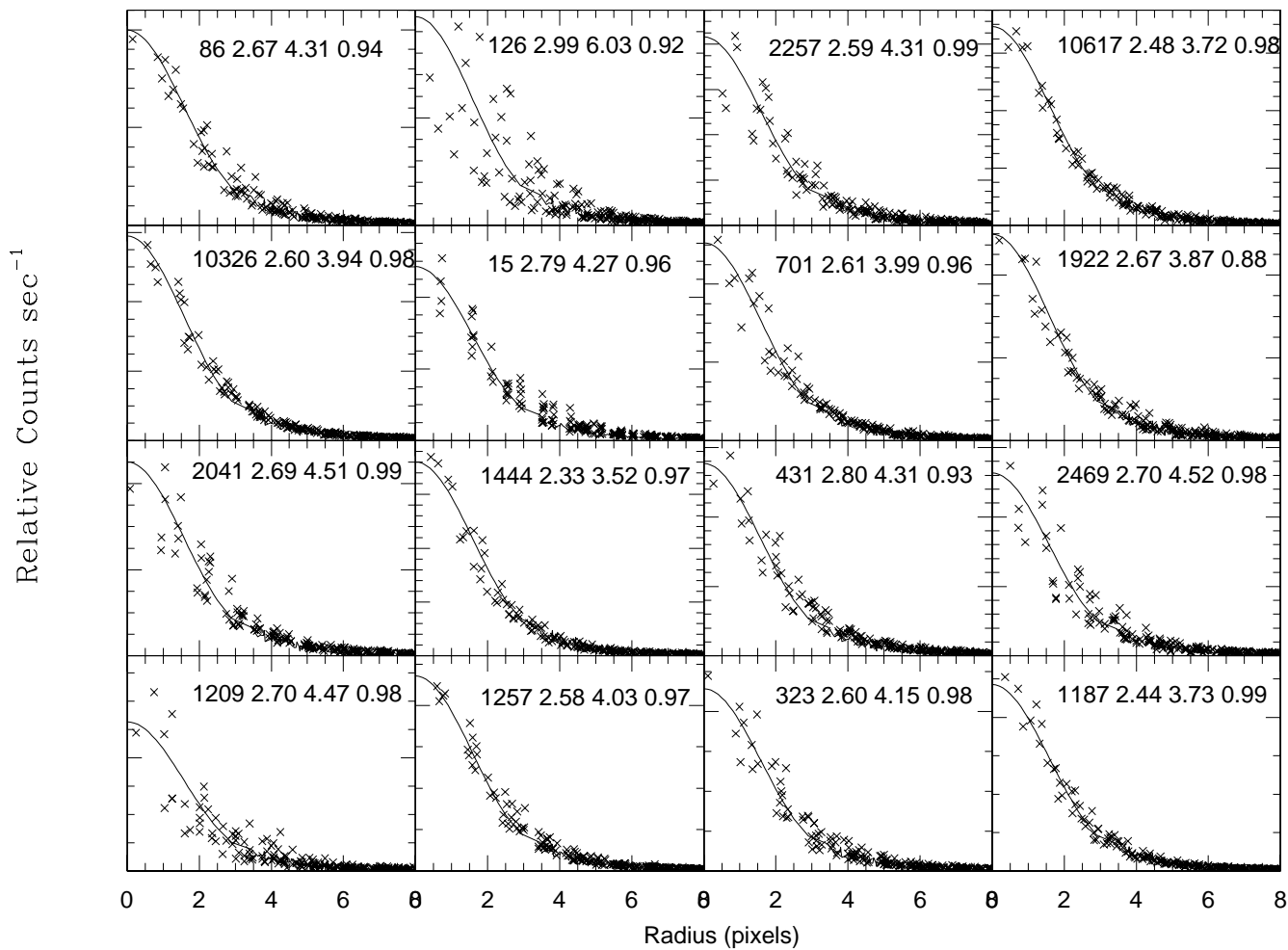}
\caption{Point spread function distributions of the objects with $V+I>21$, R50 $<$ 3 pixels and stellarity index
$>$ 0.85.The object number, R50, FWHM, and stellarity index of each object are shown in the top right corner of each
panel. The HDFS 10617 PSF is shown in each panel for comparison.}
\epsscale{1.0}
\end{figure}

\begin{figure}
\figurenum{2}
\includegraphics[angle=-90,scale=.7]{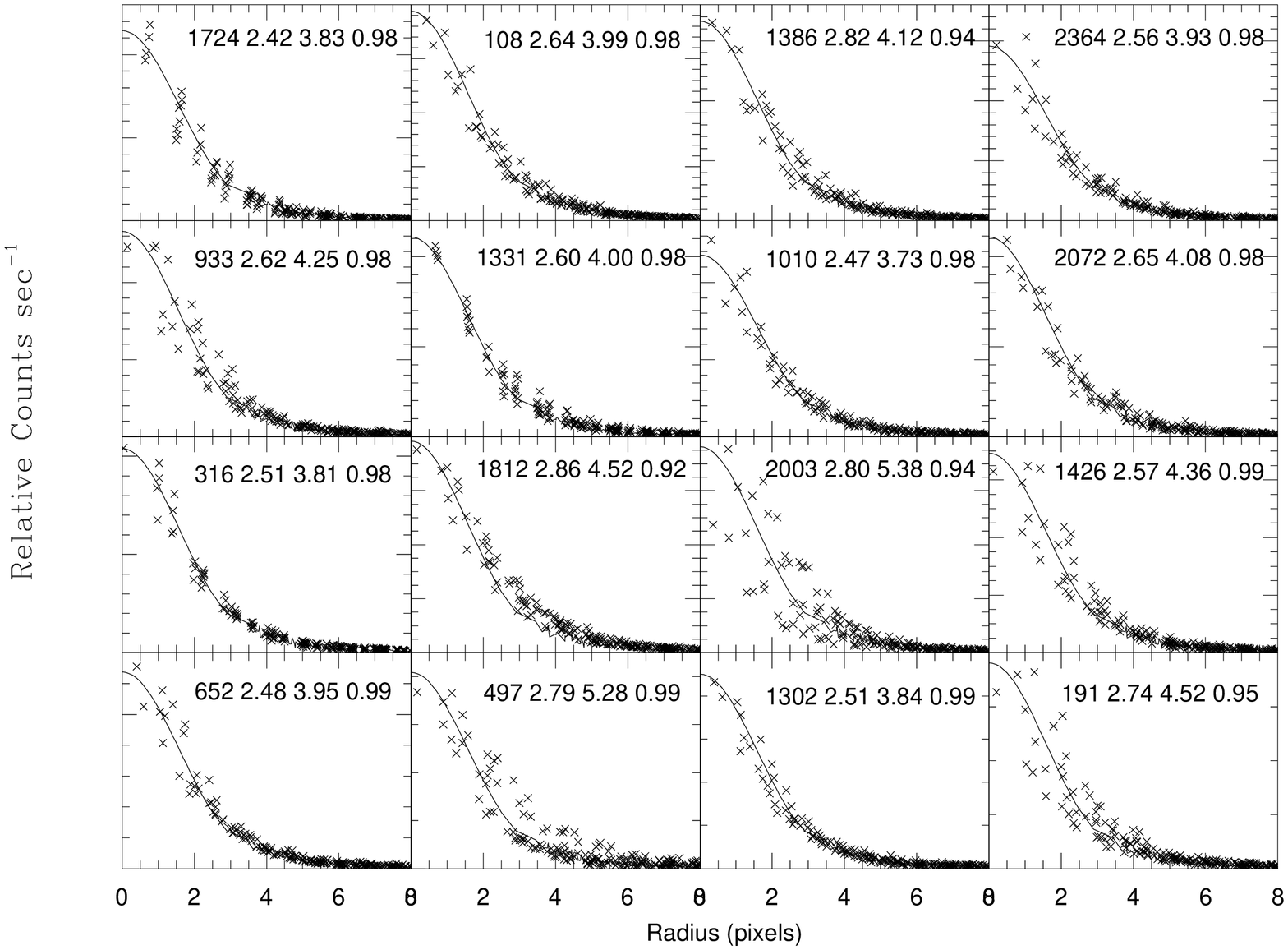}
\caption{cont.}
\end{figure}
                                                                                                                    
\begin{figure}
\figurenum{2}
\includegraphics[angle=-90,scale=.7]{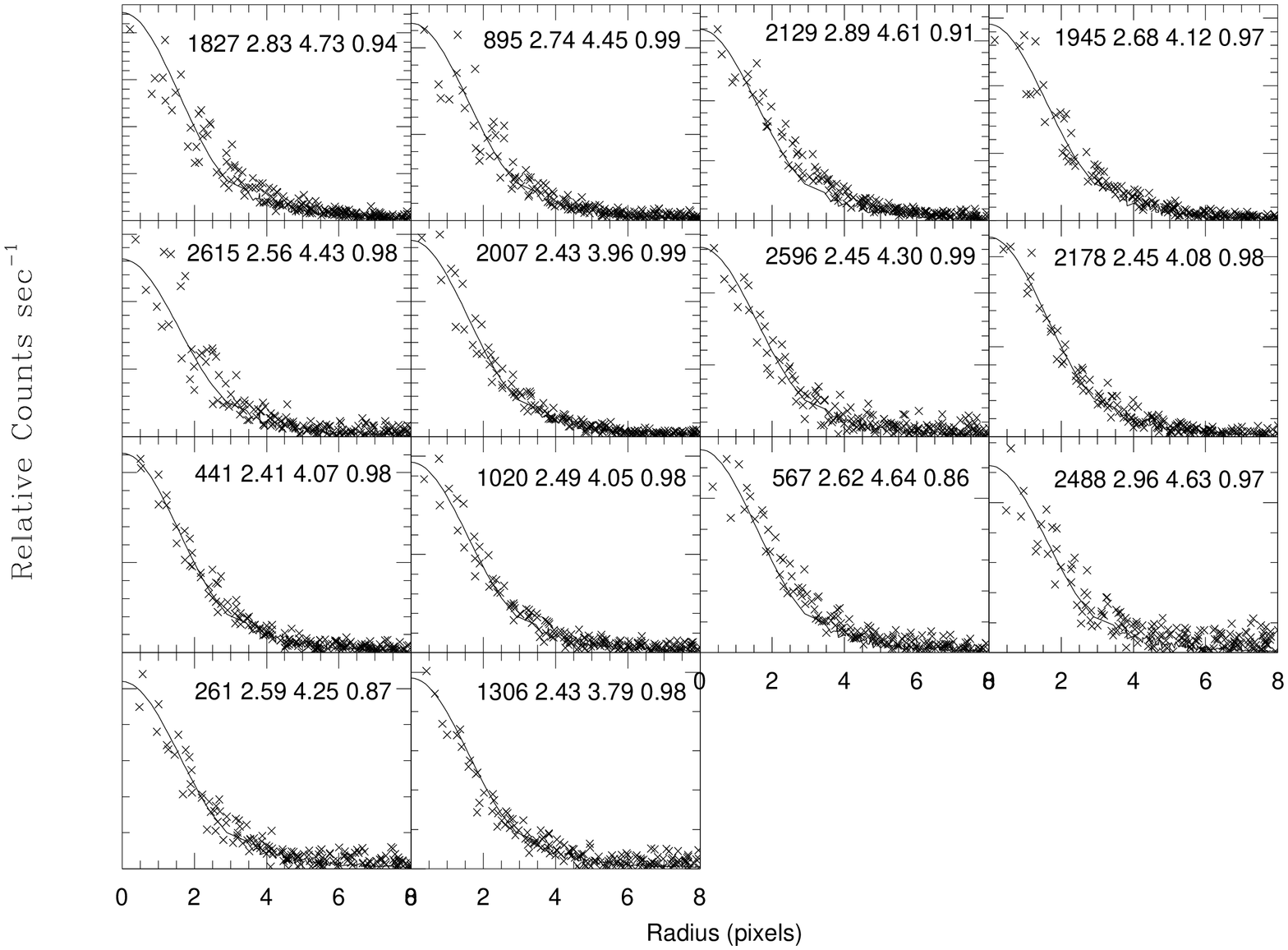}
\caption{cont.}
\end{figure}
                                                                                                                    
\begin{figure}
\figurenum{3}
\plotone{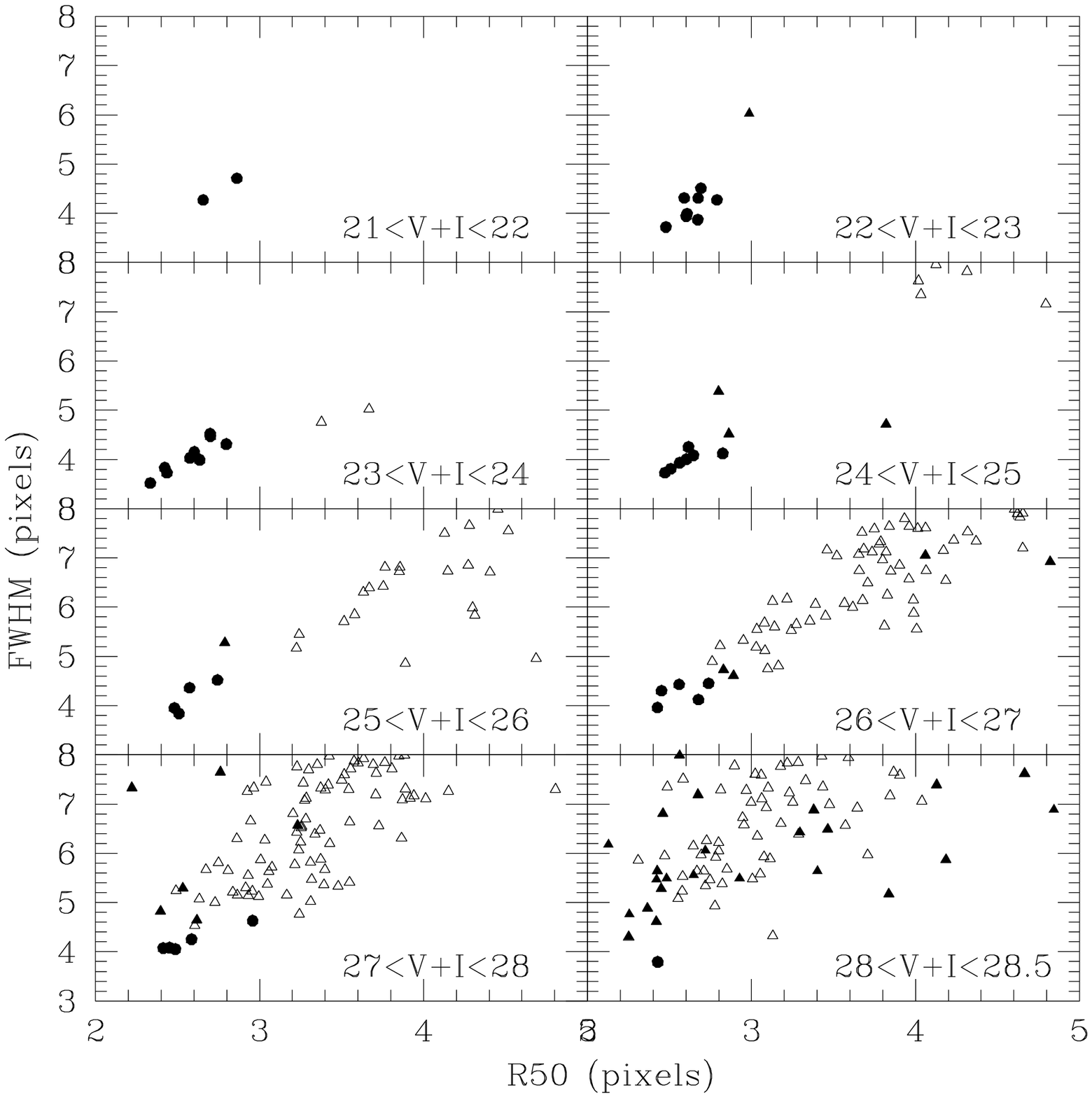}
\caption{The relation between R50 and FWHM for the HDFS objects in different magnitude ranges.
Objects with stellarity index $>$0.85 are shown as filled triangles, whereas the open triangles show
the rest of the objects in the HDFS catalog. Filled circles represent unresolved
objects identified from their empirical PSF distributions.
Point sources can be identified easily down to $V+I=27$, and the star/galaxy sequences
begin to blur for $V+I>27$.}
\end{figure}

\begin{figure}
\figurenum{4}
\plottwo{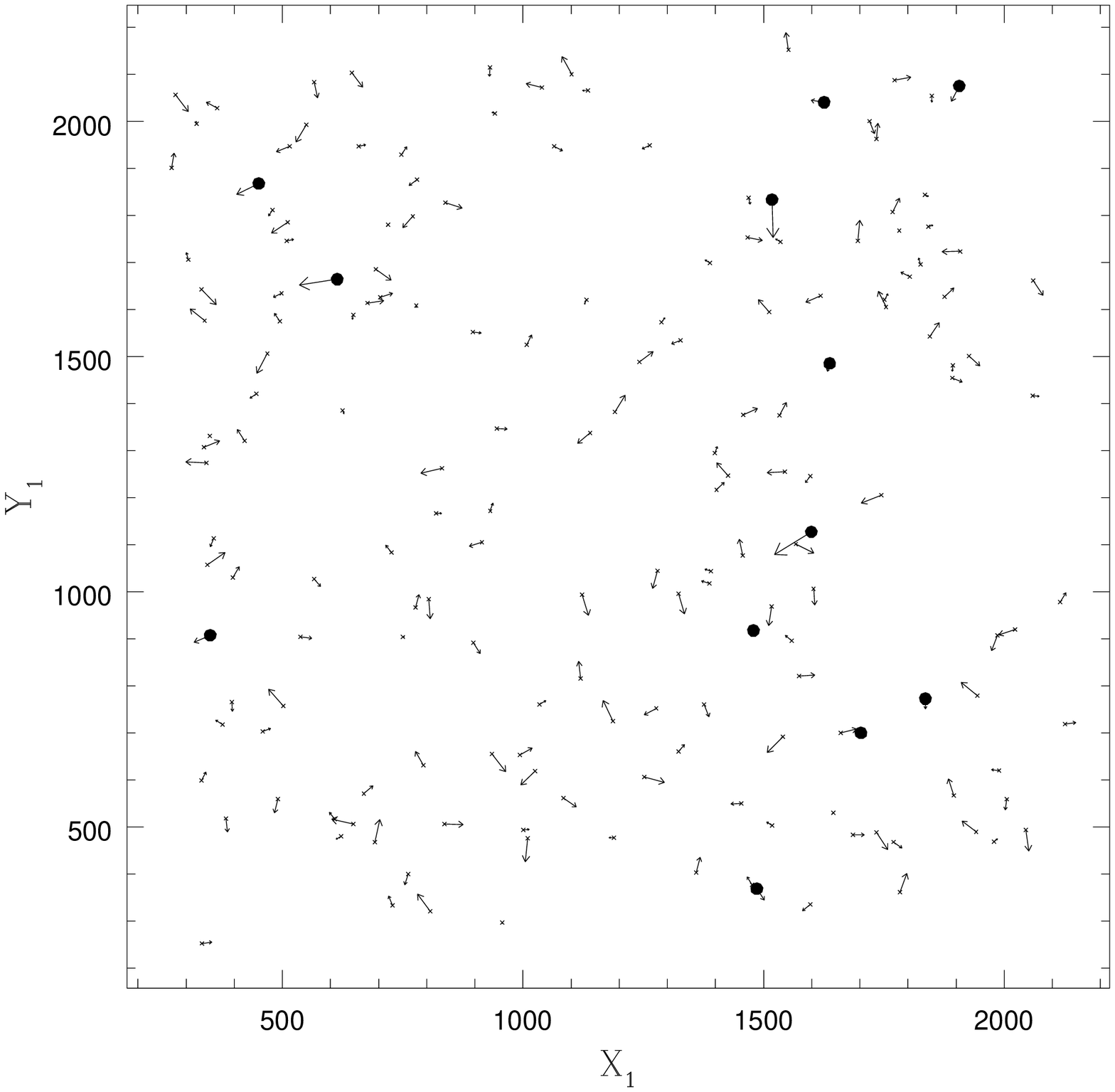}{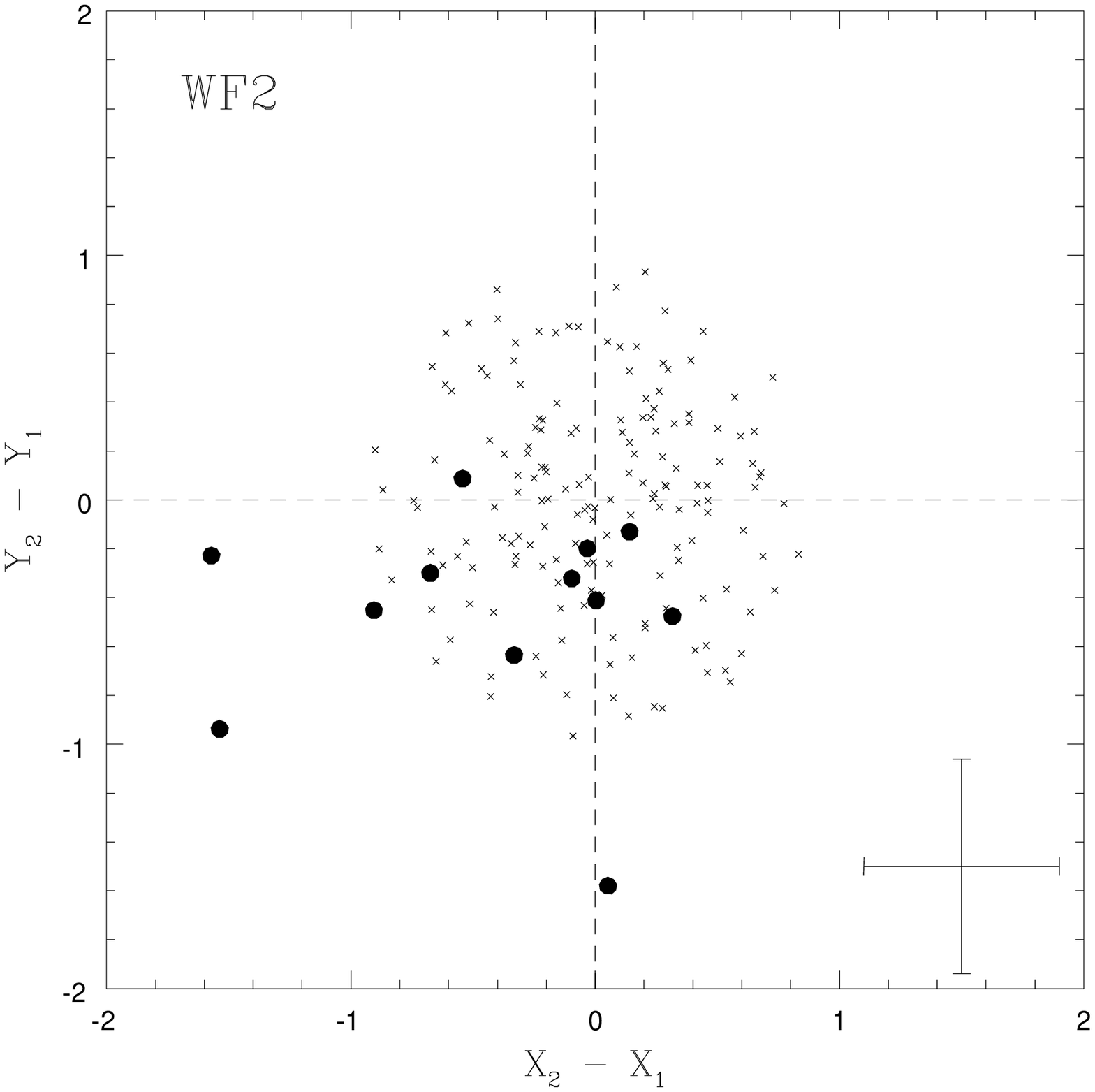}
\caption{Left panel: observed shifts (scaled by a factor of 50) of point sources (filled circles) and surrounding reference
compact objects in the HDFS WF2 chip between Epoch 1 and Epoch 2 images;
Right panel: Differences between the first-epoch coordinates (X$_1$, Y$_1$) and the transformed second-epoch
coordinates (X$_2$, Y$_2$) for the same objects.
The expected scatter caused by the transformation itself is represented by the error bar in the lower right corner of
the figure.
The axes units are HDFS pixels (0.04$\arcsec$/pixel).}
\end{figure}

\clearpage

\begin{figure}
\figurenum{5}
\plottwo{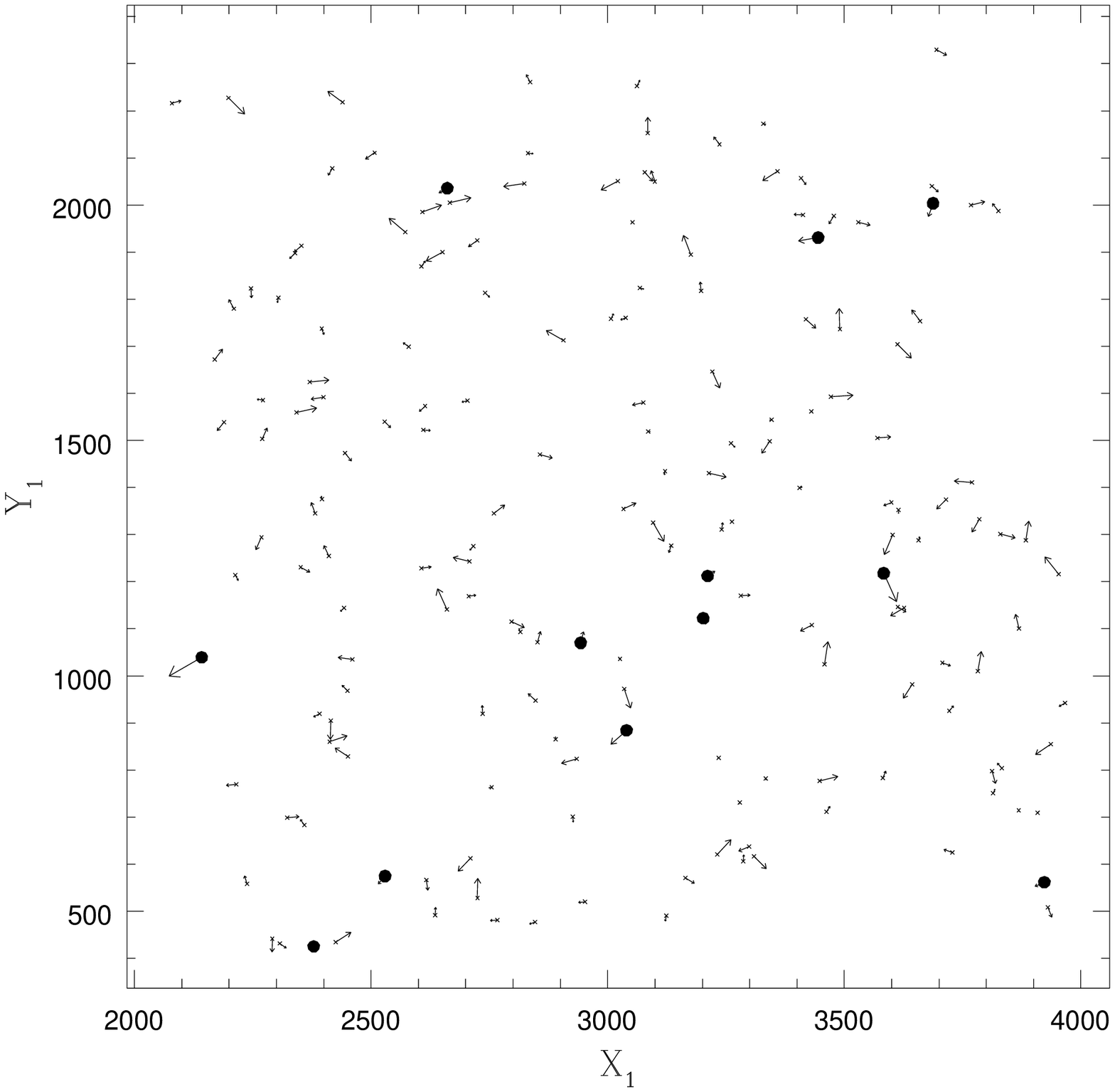}{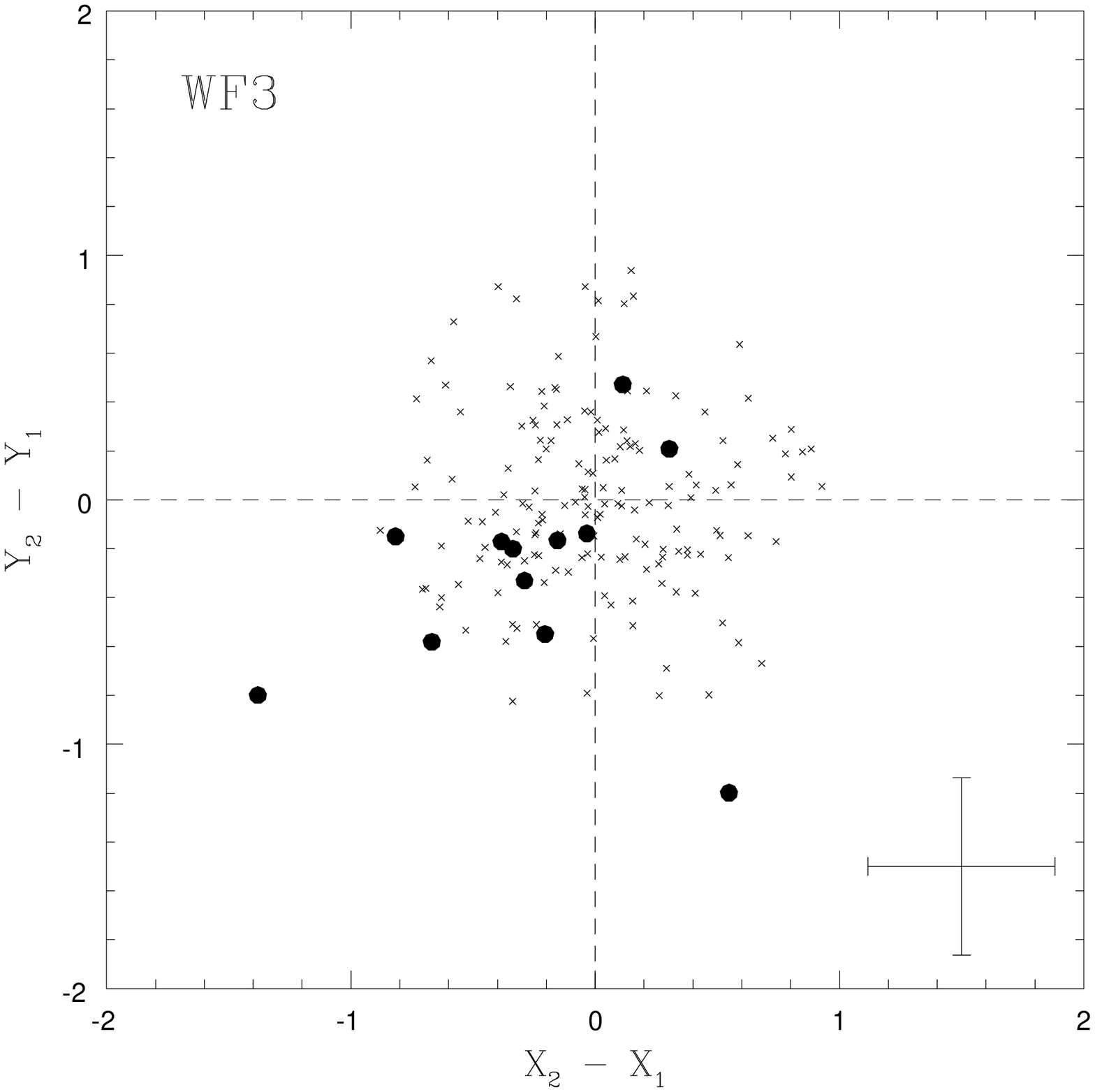}
\caption{Same as Figure 1, but for the objects detected in WF3}
\end{figure}

\clearpage

\begin{figure}
\figurenum{6}
\plottwo{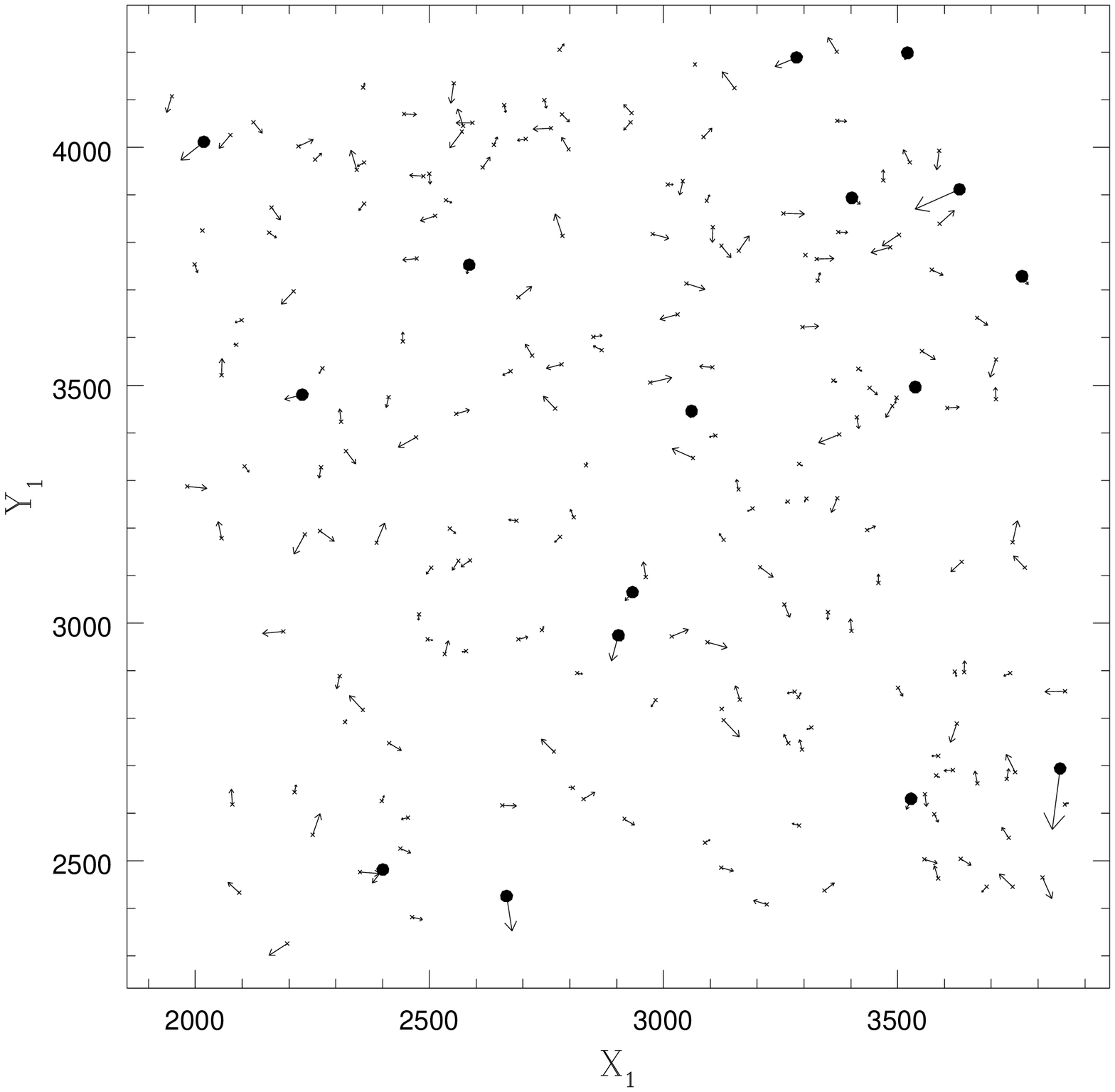}{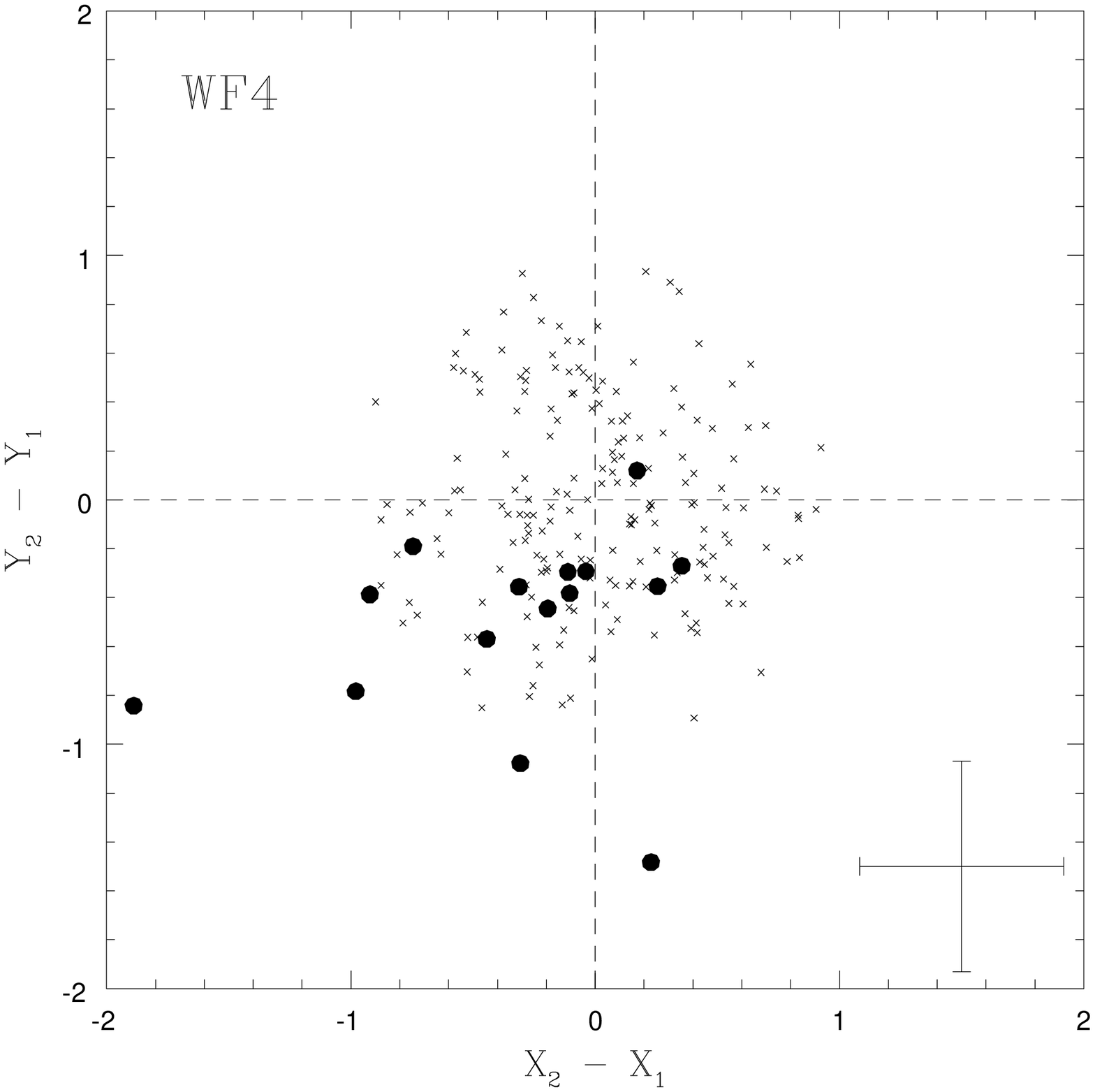}
\caption{Same as Figure 1, but for the objects detected in WF4}
\end{figure}

\clearpage

\begin{figure}
\figurenum{7}
\plotone{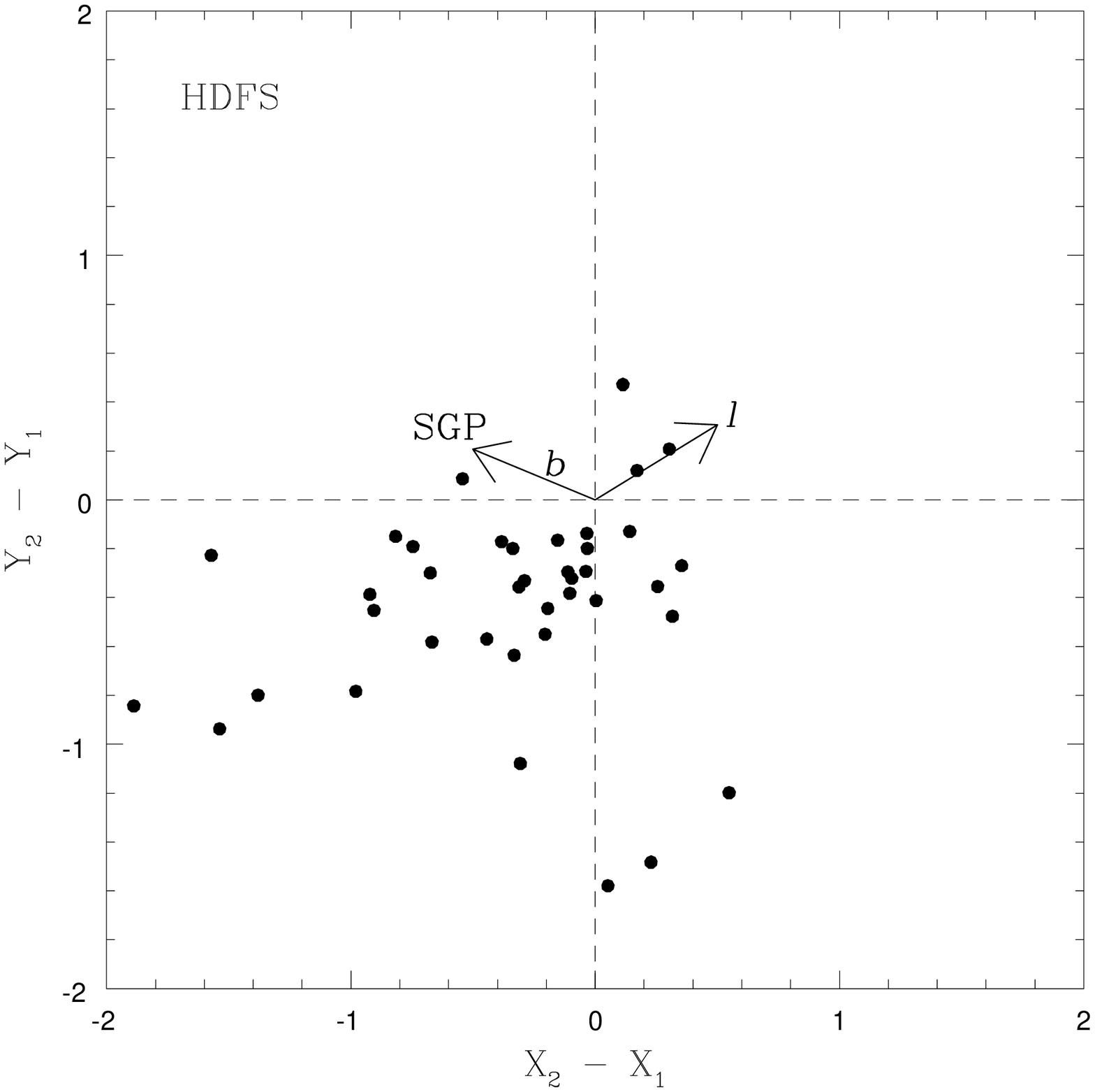}
\caption{Observed shifts of all point sources in the HDF South. The directions of increasing Galactic longitude ($\it l$) and latitude
($\it b$) and the South Galactic Pole (SGP) are also shown.}
\end{figure}

\begin{figure}
\figurenum{8}
\includegraphics[angle=-90,scale=.7]{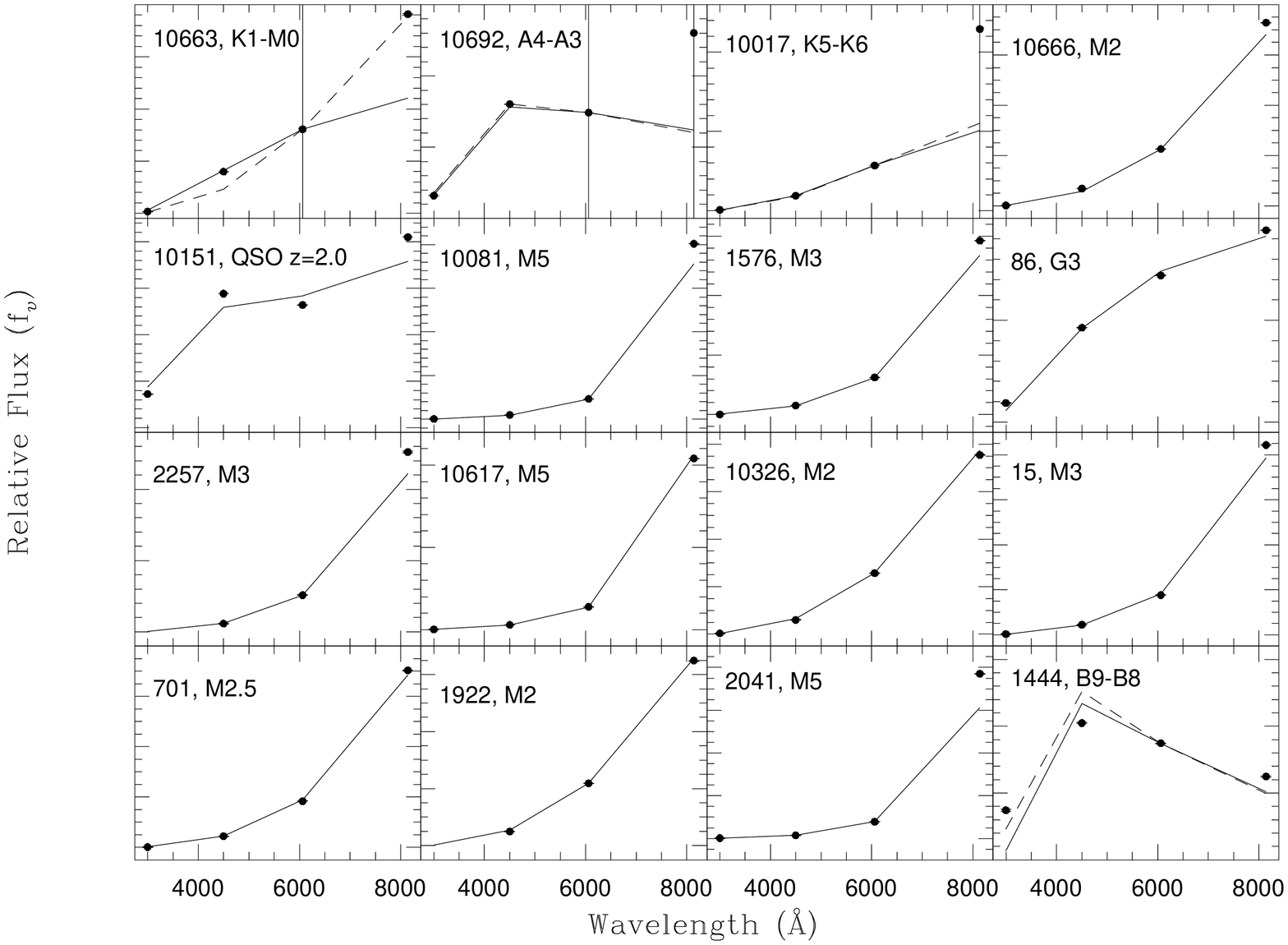}
\caption{Spectral energy distribution and best fitting template for the point sources in the HDF South.}
\end{figure}
                                                                                                                                
\begin{figure}
\figurenum{8}
\includegraphics[angle=-90,scale=.7]{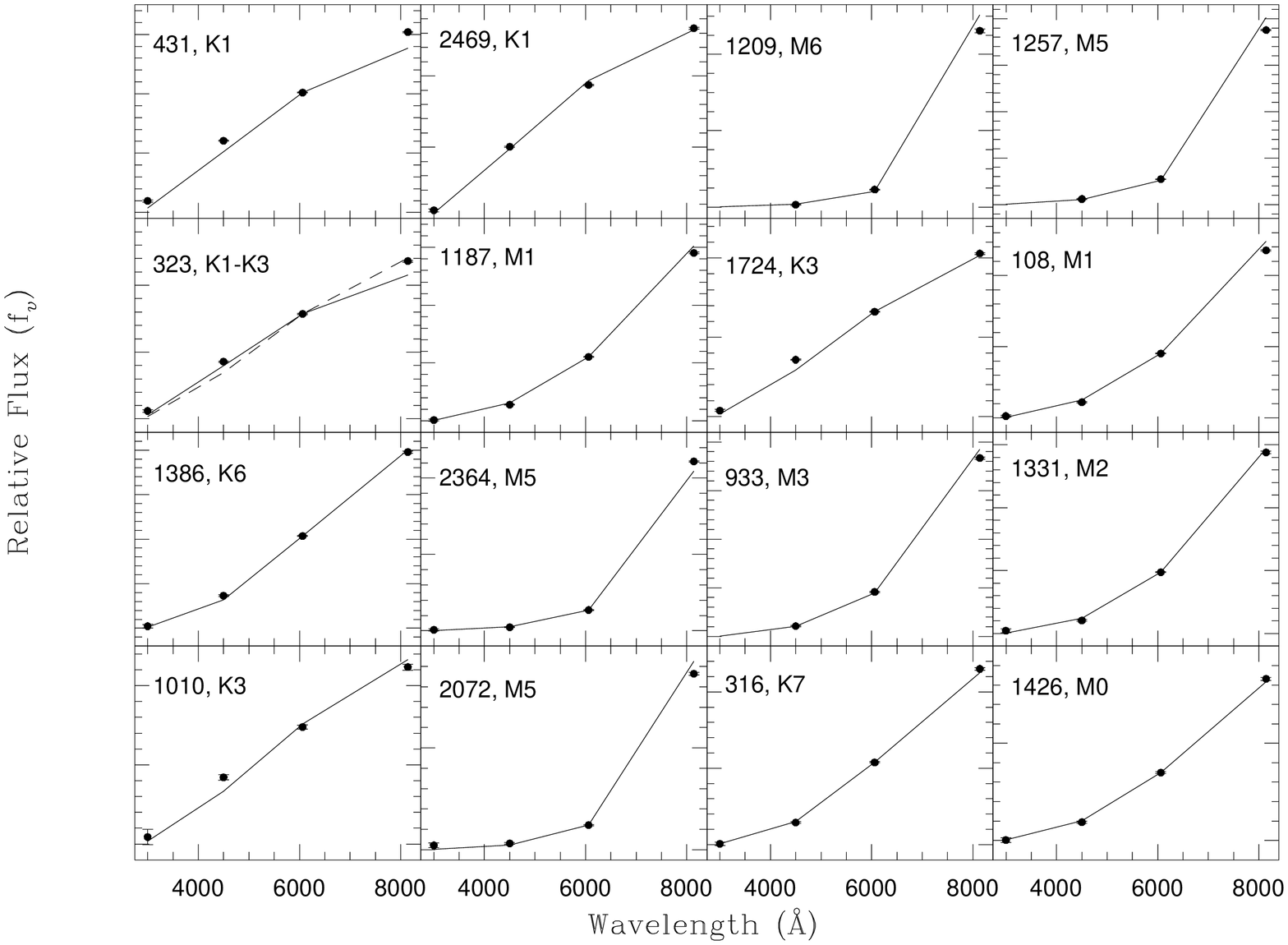}
\caption{cont.}
\end{figure}
                                                                                                                                
\begin{figure}
\figurenum{8}
\includegraphics[angle=-90,scale=.7]{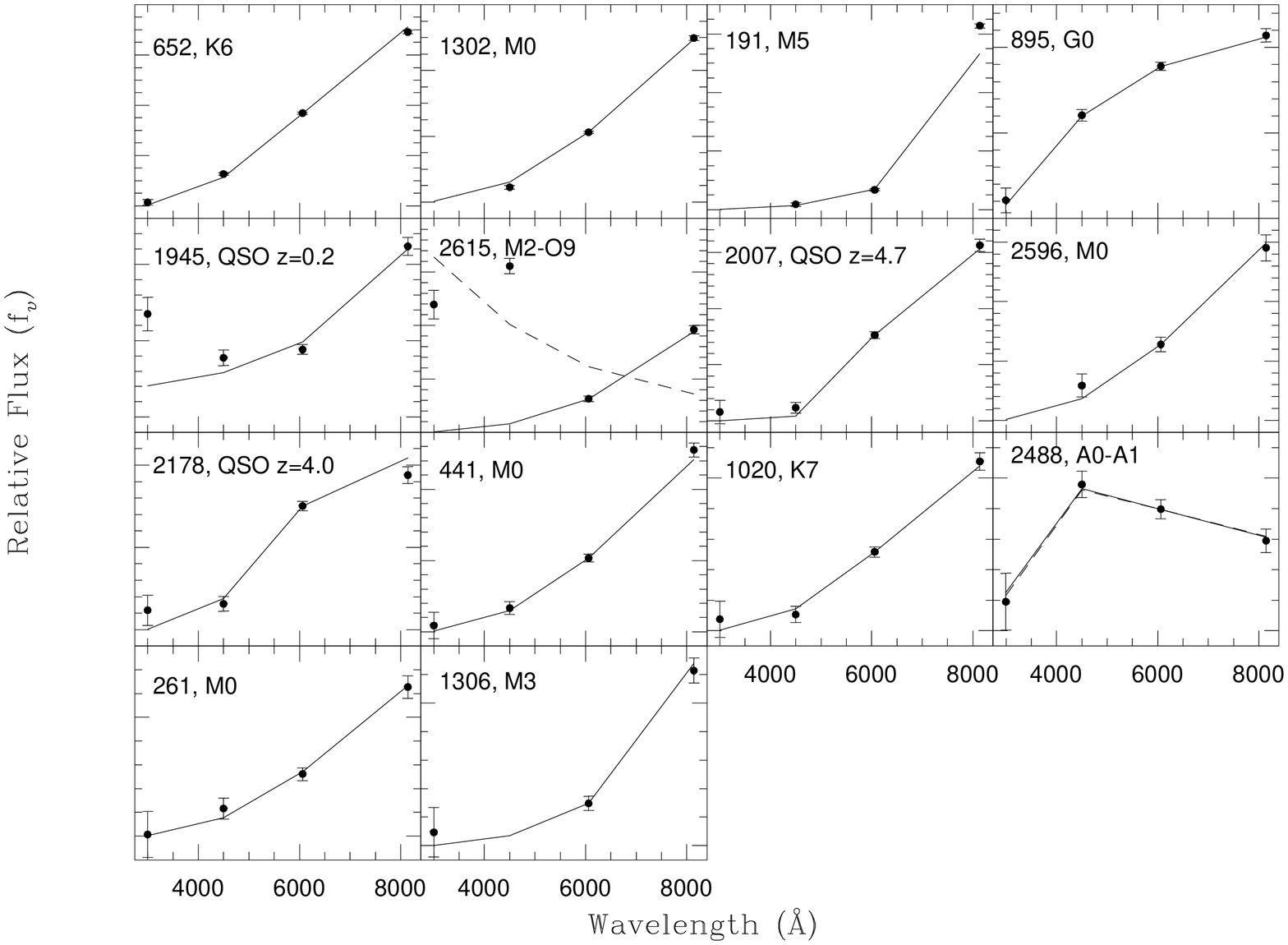}
\caption{cont.}
\end{figure}

\clearpage
\begin{figure}
\figurenum{9}
\plotone{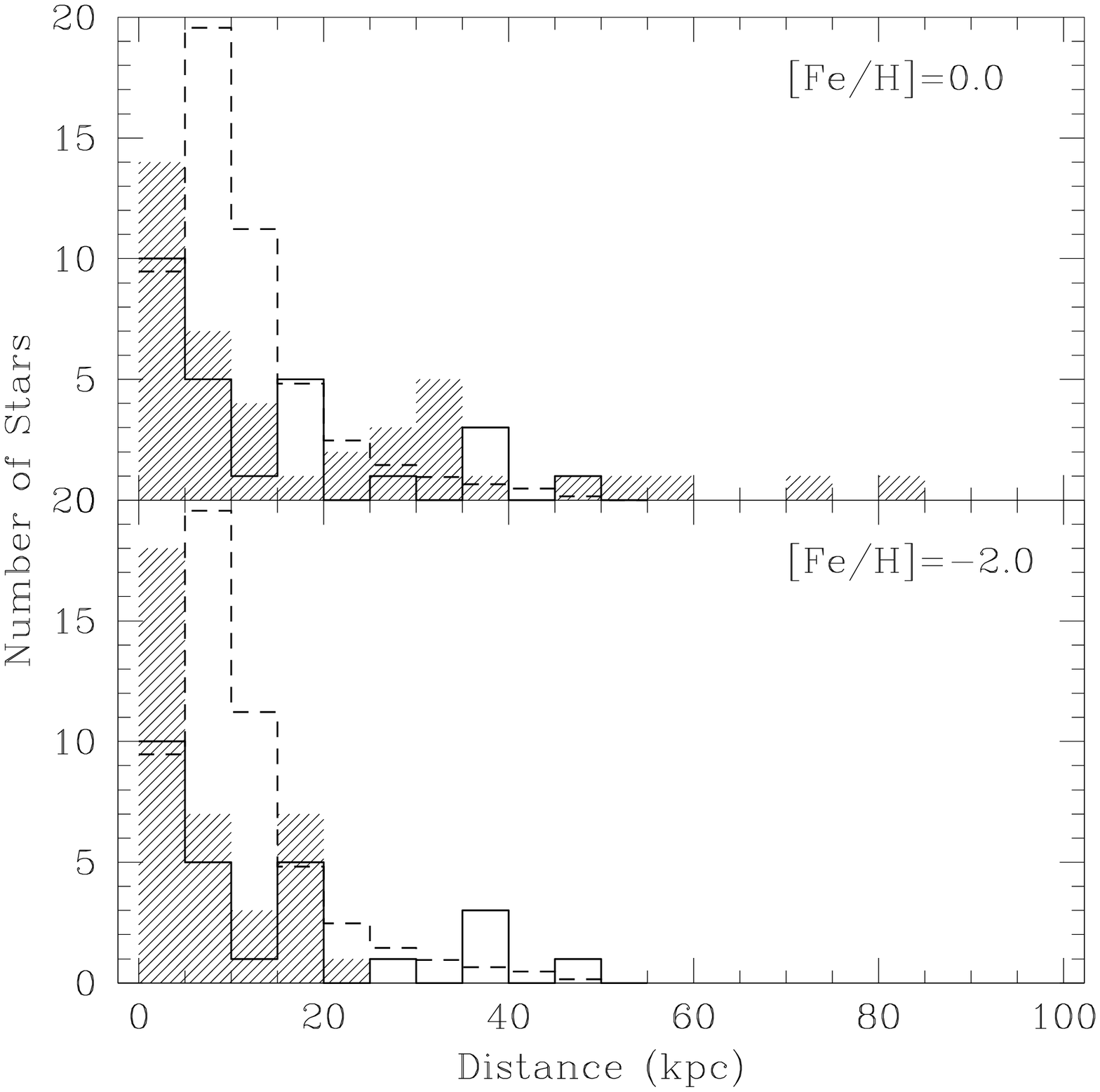}
\vspace{-0.5in}
\caption{Histogram of the number of stars observed at a given distance in the HDF South (shaded histogram) assuming solar
(top panel) or metal poor composition (bottom panel). The histogram of the number of stars observed in the HUDF (solid
line) and the predictions from the star count models (dashed-dotted line) are also shown.}
\end{figure}

\clearpage
\begin{figure}
\figurenum{10}
\plotone{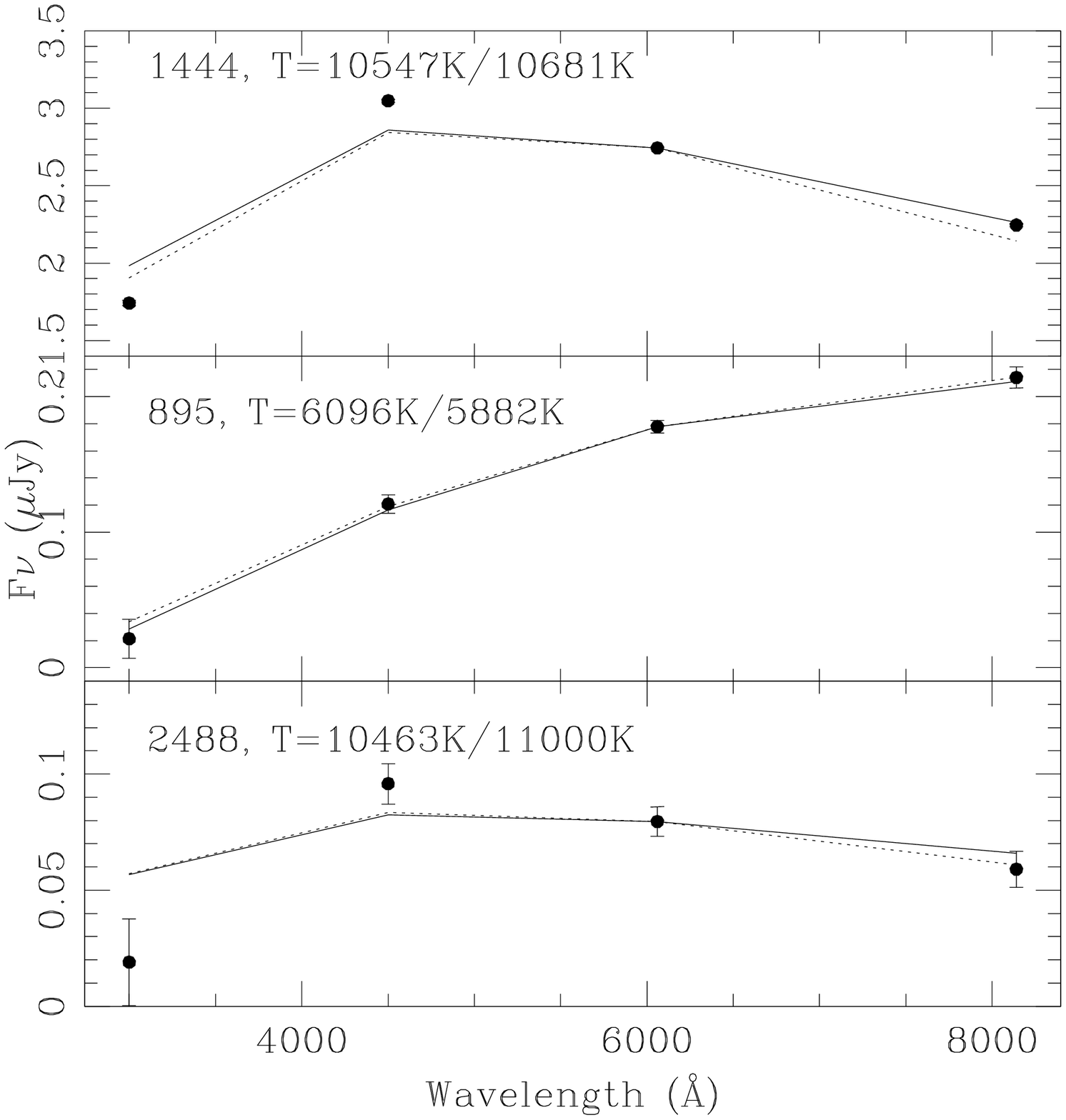}
\caption{Fits to the spectral energy distributions of HDFS 1444, 895, and 2488. Best fitting blackbody models are shown as solid lines,
whereas bestfitting pure-H atmosphere white dwarf models are shown as dotted lines. The first temperature estimate in the label
is for the blackbody and the latter one is for the pure-H white dwarf models.}
\end{figure}

\clearpage
\begin{figure}
\figurenum{11}
\plotone{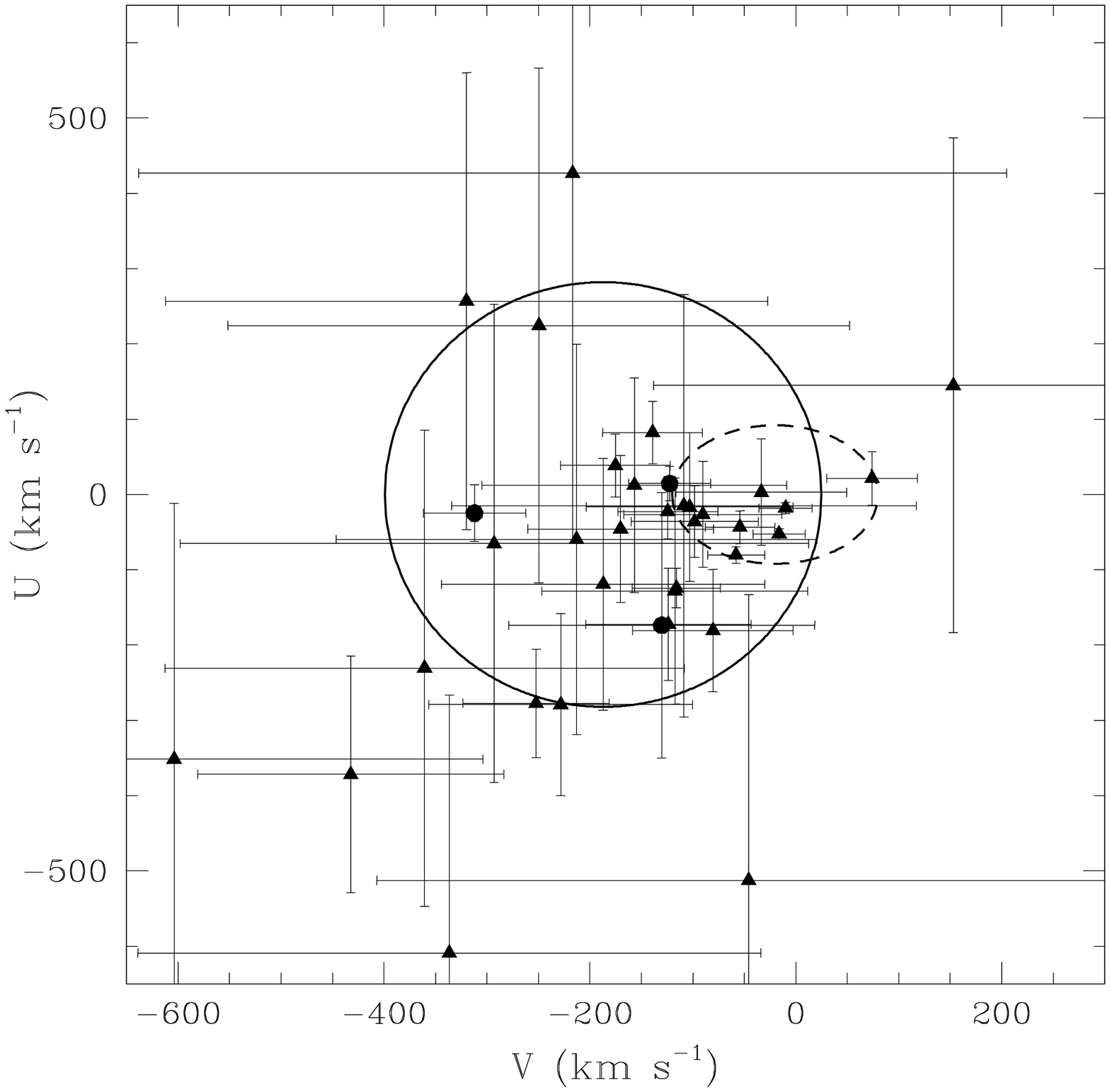}
\caption{$U$ vs. $V$ diagram for the stars (filled triangles) and the likely white dwarfs (filled circles)
in the HDF South. The
2$\sigma$ velocity ellipse of the thick disk objects is shown (dashed line) along with the 2$\sigma$ ellipse of the halo
population (solid line). The errors in the tangential velocities are larger for distant objects, therefore these objects
have also large errorbars in the $U$ vs. $V$ diagram.}
\end{figure}

\clearpage
\begin{figure}
\figurenum{12}
\plotone{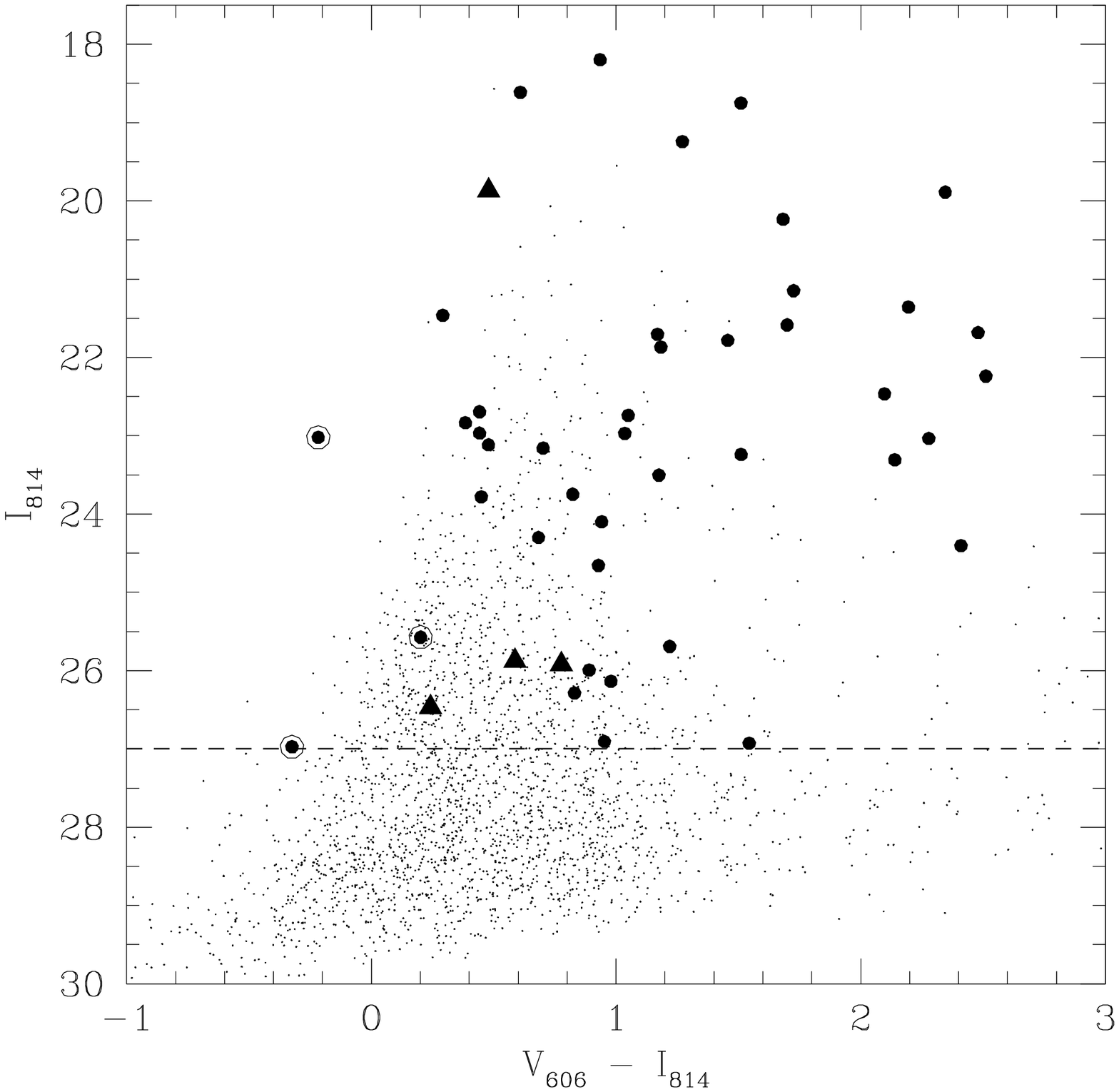}
\caption{The location of the 46 point sources in a color magnitude diagram. Filled circles and triangles represent stars
and quasars identified from spectral energy distribution fitting, respectively. White dwarf candidates are marked with
open circles, whereas the smaller dots show the rest of the objects in the HDF South catalog.}
\end{figure}

\begin{figure}
\figurenum{13}
\plotone{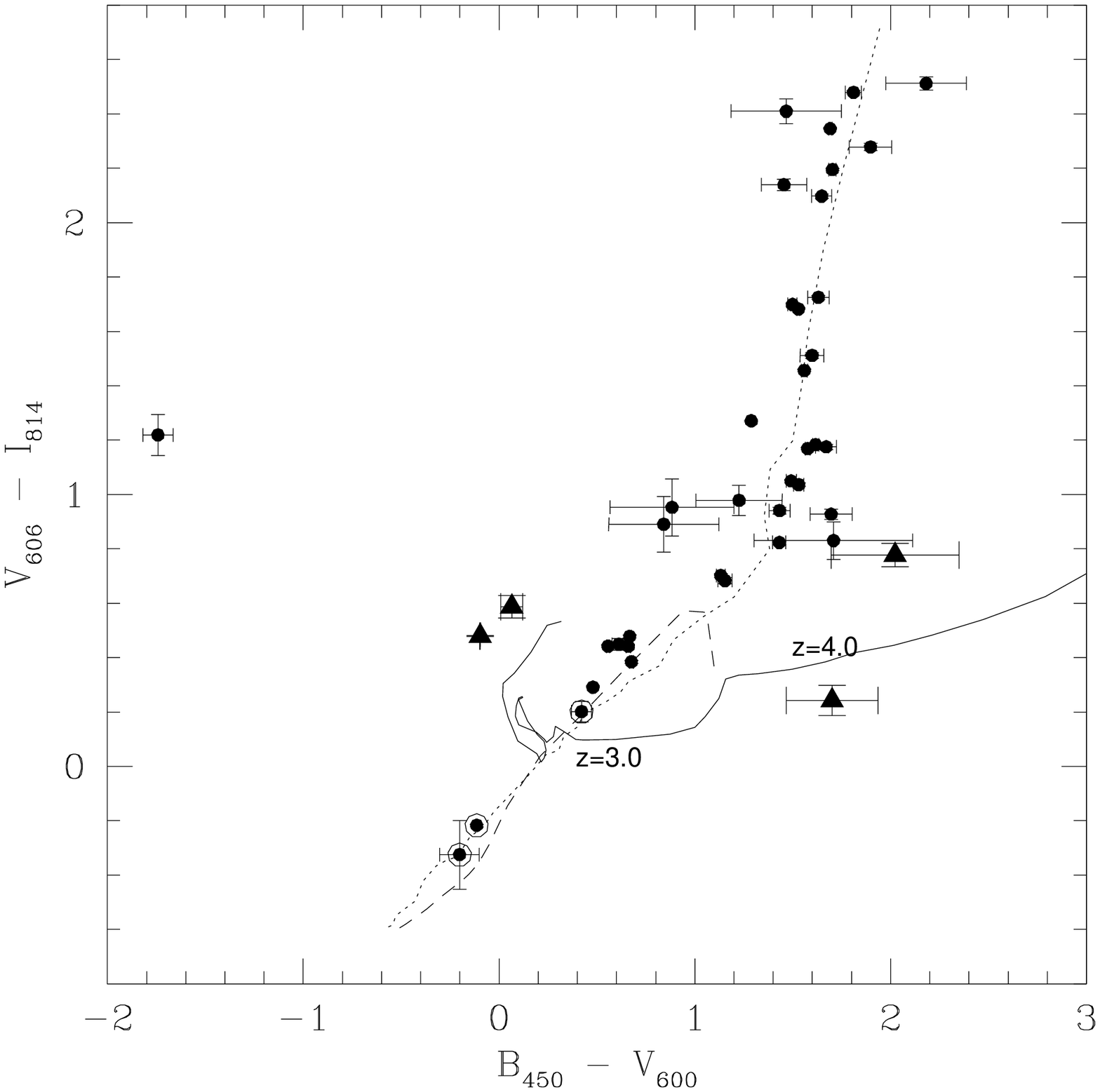}
\caption{$V_{\rm 606}-I_{\rm 775}$ vs. $B_{\rm 435}-V_{\rm 606}$ color-color diagram for the stars (circles),
the quasars (triangles) and the white dwarf candidates (open circles) in the HDF South.
Predicted tracks for main sequence stars (O5--M6; dotted line), pure-H white dwarfs ($log$ g=8, $T_{\rm eff}=$ 60000--3000K;
dashed line), and quasars (solid line) are also shown.}
\end{figure}

\end{document}